\documentclass{jpsj3}
\usepackage{graphicx}
\usepackage{latexsym}
\title{Oscillations in Spurious States of the Associative Memory Model with Synaptic Depression}
\author{\name{Shin \surname{Murata}}$^1$, \name{Yosuke \surname{Otsubo}}$^{1,2}$, \name{Kenji \surname{Nagata}}$^1$, and \name{Masato \surname{Okada}}$^{1,3}$\thanks{E-mail:okada@k.u-tokyo.ac.jp}}
 \inst{$^1$Graduate School of Frontier Science, The University of Tokyo, 5-1-5 Kashiwanoha, Kashiwa-shi, Chiba-ken 277-8561, Japan \\
 $^2$Research Fellow of the Japan Society for the Promotion of Science\\
 $^3$RIKEN Brain Science Institute, 2-1 Hirosawa, Wako-shi, Saitama-ken 351-0198, Japan}
 \abst{
The associative memory model is a typical neural network model, which can store discretely distributed fixed-point attractors as memory patterns.
When the network stores the memory patterns extensively, however, the model has other attractors besides the memory patterns. 
These attractors are called spurious memories. 
Both spurious states and memory states are equilibrium, so there is little difference between their dynamics. 
Recent physiological experiments have shown that short-term dynamic synapse called synaptic depression decreases its transmission efficacy to postsynaptic neurons according to the activities of presynaptic neurons. 
Previous studies have shown that synaptic depression induces oscillation in the network and decreases the storage capacity at finite temperature. 
How synaptic depression affects spurious states, however, is still unclear. 
We investigate the effect of synaptic depression on spurious states through Monte Carlo simulation. 
The results demonstrate that synaptic depression does not affect the memory states but mainly destabilizes the spurious states and induces the periodic oscillations.}
  \kword{	
  associative memory model, synaptic depression, spurious state, principal component analysis
  }
  \begin{document}
  \maketitle
 \section{Introduction} 
 The associative memory model is a typical neural network model.
 The network in this model has discretely distributed fixed-point attractors,
 and it can store these point attractors as memory patterns by using Hebbian rule\cite{Nakano1972,Hopfield1982,Okada1996}.
 Given an initial state near a memory pattern, the network generally converges on the memory pattern.
 The associative memory model has other attractors, however, besides the stored memory patterns in extensive loading cases, which means that the number of memory patterns is on the order of the number of neurons.
 These other attractors are called spurious memories.
 A state in which the network converges on a memory pattern is called a memory state, while a state in which the network converges on a spurious memory is called a spurious state.
 There is little difference in network dynamics between memory states and spurious states, because both states are equilibrium.
 This is a serious problem in applying the associative memory model as an information processing model.
  
 Recent physiological experiments have shown that the efficacy of synaptic transmission changes in a short period of time.
 For example, dynamic synapses can decrease their transmission efficacy to postsynaptic neurons according to the activities of presynaptic neurons\cite{Thomson1994,Abbott1997,Markram1996,Tsodyks1997}. 
 This phenomenon is called synaptic depression and is a kind of short-term synaptic plasticity.
 Previous studies have revealed that synaptic depression destabilizes attractors and induces switching phenomena among them when the number of memory patterns is much smaller than the number of neurons, i.e.,\ in finite loading cases\cite{Pantic2002,Otsubo2010}.
 However, the network has many attractors including spurious memories in extensive loading cases.
 Effects of synaptic depression on these attractors have not been investigated in detail.
 Synaptic depression may induce difference between memory states and spurious states.

 In this study, we apply Monte Carlo simulation in order to reveal effects of synaptic depression on attractors including spurious memories.
 As a result, we find that synaptic depression does not affect memory states but mainly destabilizes spurious states and induces periodic oscillations.
 This result suggests that incorporating synaptic depression would improve the performance of the associative memory model as an information processing model.

 This paper consists of four sections, including this Introduction.
 In Sect. 2 we describe the associative memory model with synaptic depression.
 Section 3 describes and analyzes the simulation results.
 Finally, in Sect. 4 we summarize the results.
 
 \section{Model}
 In this section, we introduce an associative memory model with synaptic depression, consisting of $N$ fully connected McCulloch-Pitts neurons.
 If the $i$-th neuron fires at time $t$, its state is represented by $s_i(t)=1$; otherwise, $s_i(t)=0$.
 The network state at time $t$ is denoted by a variable  $\boldsymbol{s}(t) = \left(s_1(t),s_2(t),\cdots,s_N(t)\right)$.
 Each neuron is updated with the following probability:
 \begin{equation}
  \mathrm{Prob}\left[s_i\left(t+1\right)=1\right]=1-\mathrm{Prob}\left[s_i\left(t+1\right)=0\right]=F\left(h_i\left(t\right)\right),\label{eq:renew}
 \end{equation}
 \begin{equation}
  h_i\left(t\right)=\sum_{j \neq i}^N J_{ij}\left(t\right)s_j\left(t\right).
 \end{equation}
 Here, $h_i(t)$ represents an internal potential in the $i$-th neuron at time $t$, and $F\left(\cdot\right)$ is a function taking continuous values on $[0,1]$.
 We consider an inverse temperature $\beta=1/T$ and then we define the function $F\left(\cdot\right)$ as
 \begin{equation}
  F\left(x\right)=\frac{1}{2}\left(1+\tanh\left(\beta x\right)\right).
 \end{equation}
 The strength of a synaptic connection between the $i$-th and $j$-th neurons is denoted by $J_{ij}\left(t\right)$ and given by the following:
 \begin{eqnarray}
  J_{ij}\left(t\right)&=&\tilde{J}_{ij}x_j(t),\\
  x_j(t+1)&=&x_j(t)+\frac{1-x_j(t)}{\tau}-U_{SE} x_j(t)s_j(t),
 \end{eqnarray}
 where $x_j(t)$ is the efficacy of synaptic transmission and takes continuous values in the range $0<x_j(t)\leq 1$\cite{Tsodyks1997,Pantic2002}.
 Here, $U_{SE}$ represents the fraction of released neurotransmitter in the absence of depression, and $\tau$ is a time constant for the recovery process.
 The absolute strength of synaptic connection, $\tilde{J}_{ij}$, is given by the following Hebbian rule:
 \begin{equation}
  \tilde{J}_{ij}=\frac{1}{N}\sum_{\mu=1}^{p}\xi_i^\mu\xi_j^\mu.
 \end{equation}
 Here, $p$ corresponds to the number of memory patterns stored in the network.
 The ratio of the number of neurons, $N$, to the number of memory patterns, $p$, is called the loading ratio $\alpha$, i.e.,\ $\alpha = p/N$.  
 The stored memory pattern $\boldsymbol{\xi}^\mu = \left(\xi_1^\mu,\xi_2^\mu,\cdots,\xi_N^\mu\right)$ is stochastically provided by the following:
 \begin{equation}
  \mathrm{Prob}\left[\xi^\mu_i=\pm 1\right]=\frac{1}{2}.
 \end{equation}
 
\section{Results}
 In this section, we show the results obtained from Monte Carlo simulation with the model defined in Sect. 2.
 In order to describe macroscopic state of a network, we define an overlap between the $\mu$-th memory pattern $\boldsymbol{\xi}^\mu$ and the network state $\boldsymbol{s}(t)$ as
 \begin{equation}
  M^\mu(t)=\frac{1}{N}\sum_{i=1}^N\xi_i^\mu\left(2s_i(t)-1\right).
 \end{equation}
 The overlap $M^\mu(t)$ corresponds to the direction cosine between the $\mu$-th memory pattern and the state of the network.
 When the state $\boldsymbol{s}(t)$ completely matches the memory pattern $\boldsymbol{\xi}^\mu$, the overlap $M^\mu(t)$ takes the value $1$. 
 On the other hand, when $\boldsymbol{s}(t)$ is orthogonal to $\boldsymbol{\xi}^\mu$, then $M^\mu(t)$ takes the value $0$.
 The initial network state $\boldsymbol{s}(0)$ is given by the following probability:
 \begin{equation}
  \mathrm{Prob}\left[s_i(0)=1\right]=1-\mathrm{Prob}\left[s_i(0)=0\right]=\frac{1+M^\mu_0\xi_i^\mu}{2}.
 \end{equation}
 The expectation of the overlap $M^\mu(0)$ provided in this way becomes $M^\mu_0$ in the limit of $N \to \infty$. Therefore, $M^\mu_0$ corresponds to the initial overlap.

 \begin{figure}[b]
  \begin{tabular}{rcrc}
   (a)&
   \includegraphics[height=54mm]{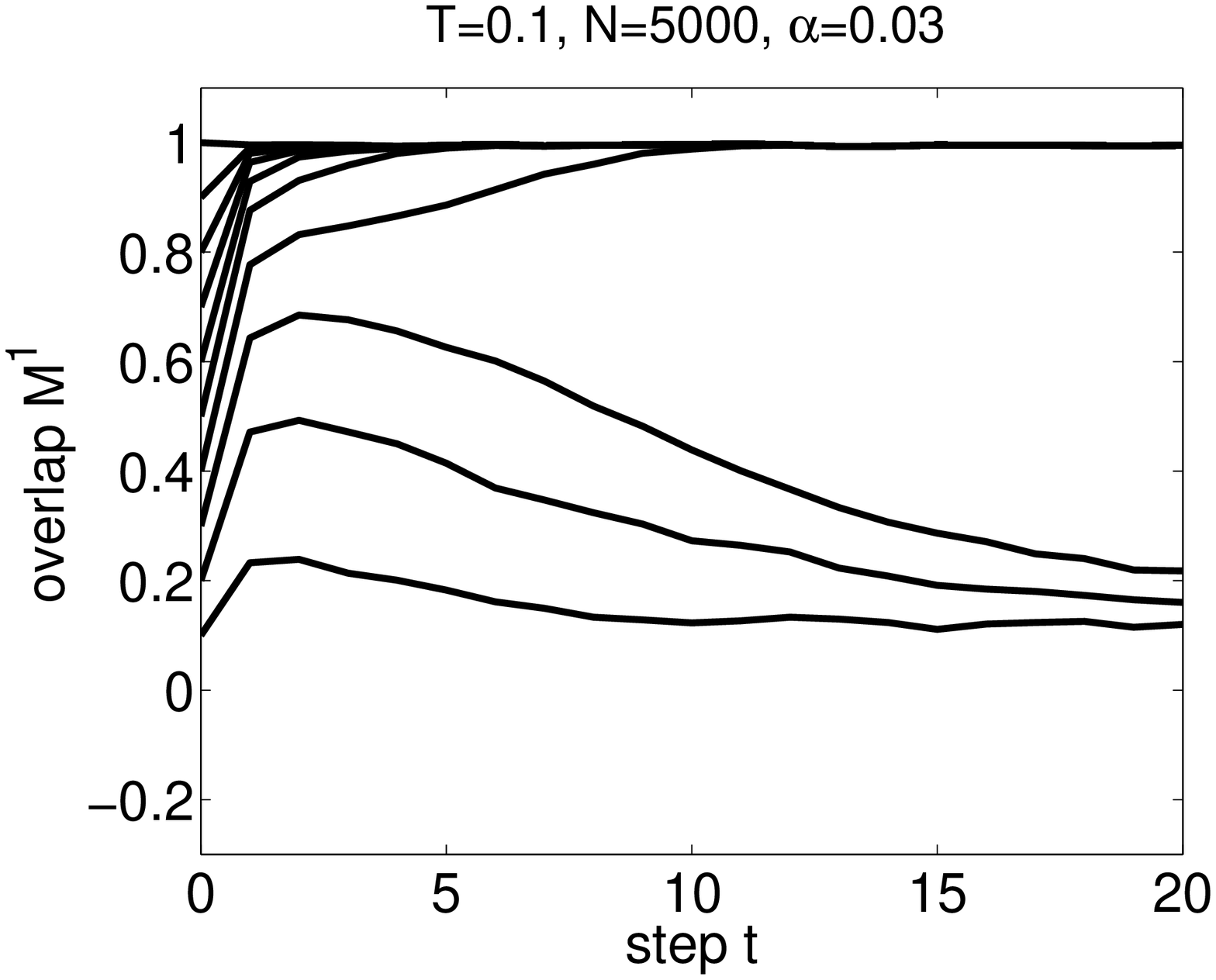}  &
   (b)&
   \includegraphics[height=54mm]{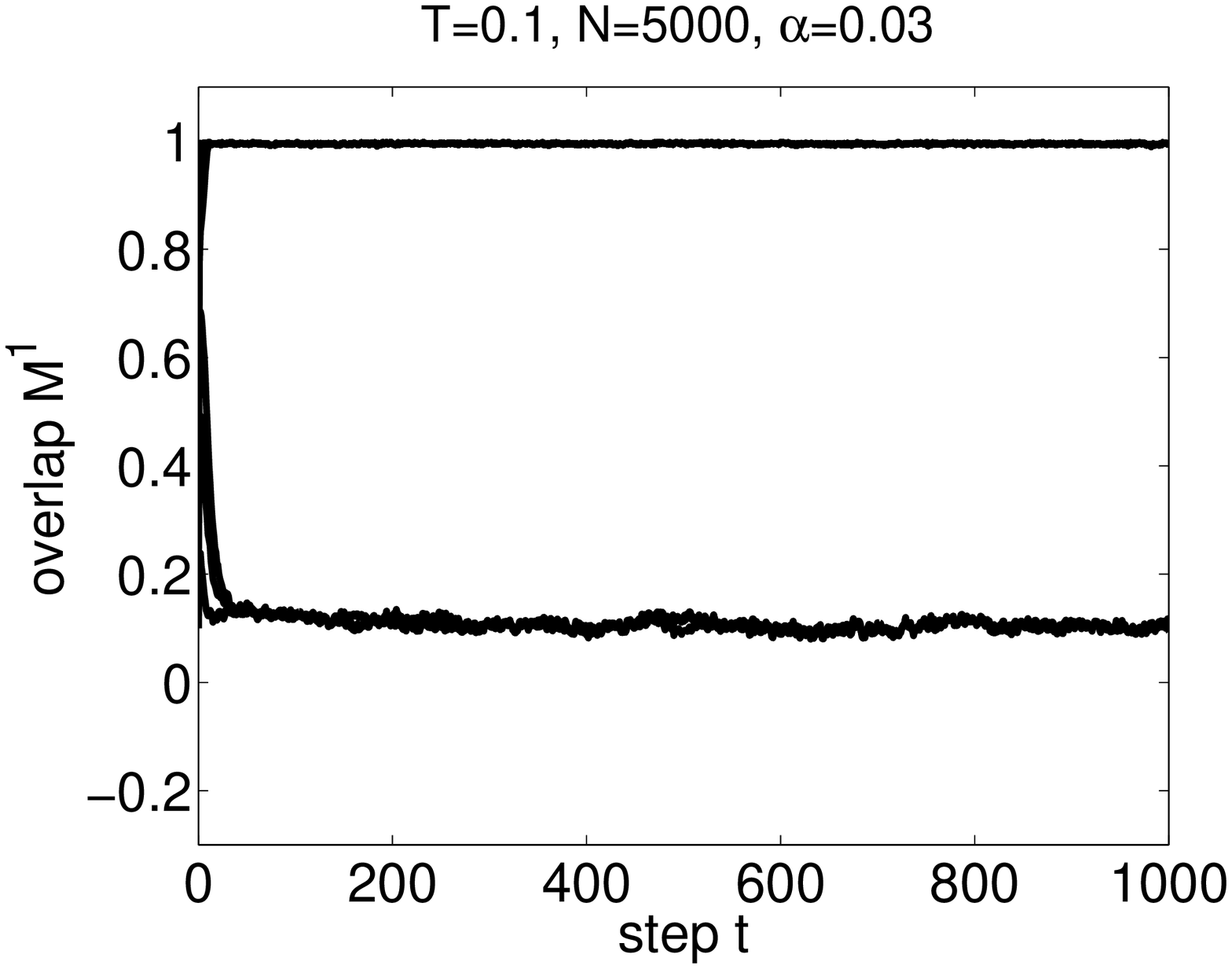}\\
   (c)&
   \includegraphics[height=54mm]{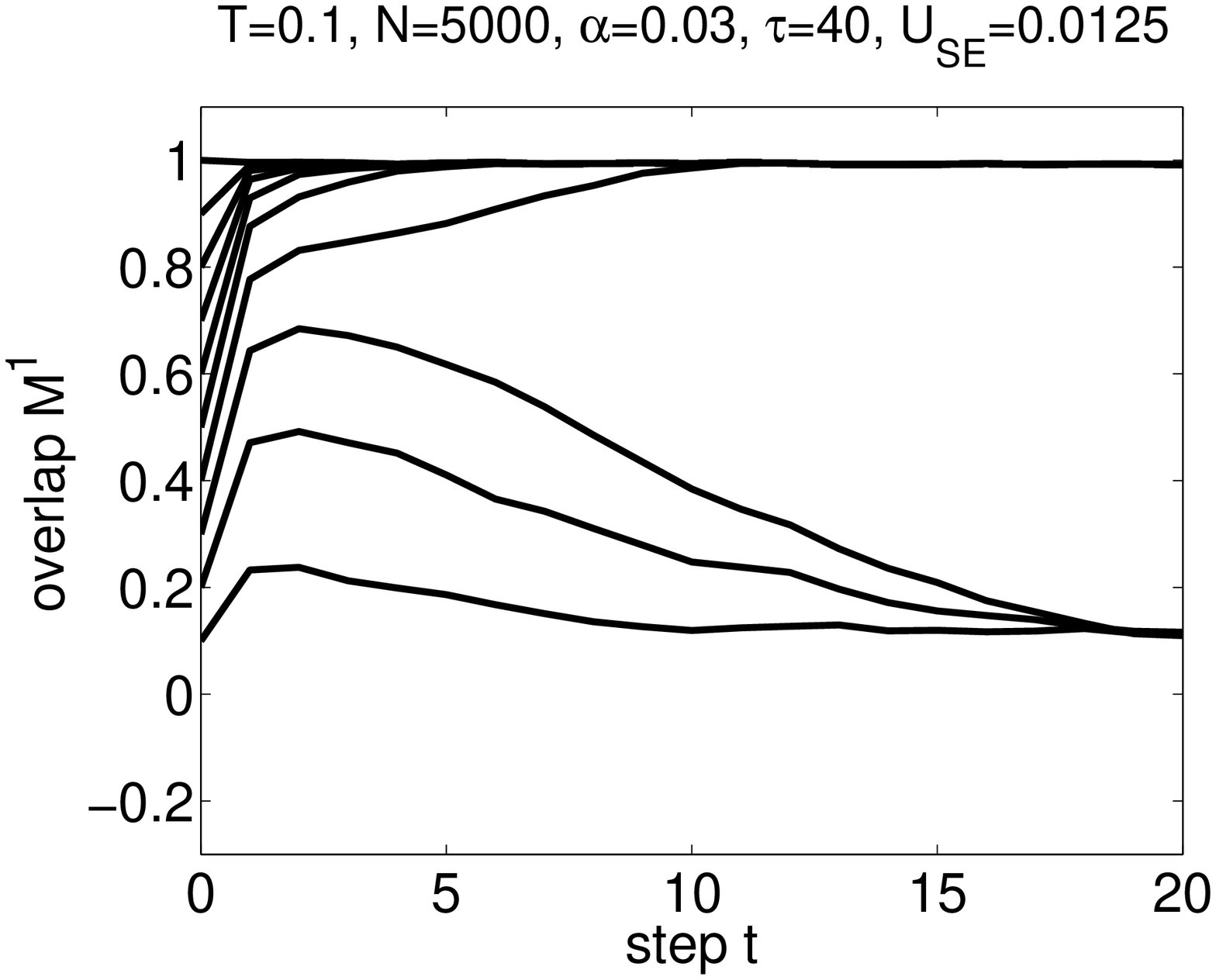}&
   (d)&
   \includegraphics[height=54mm]{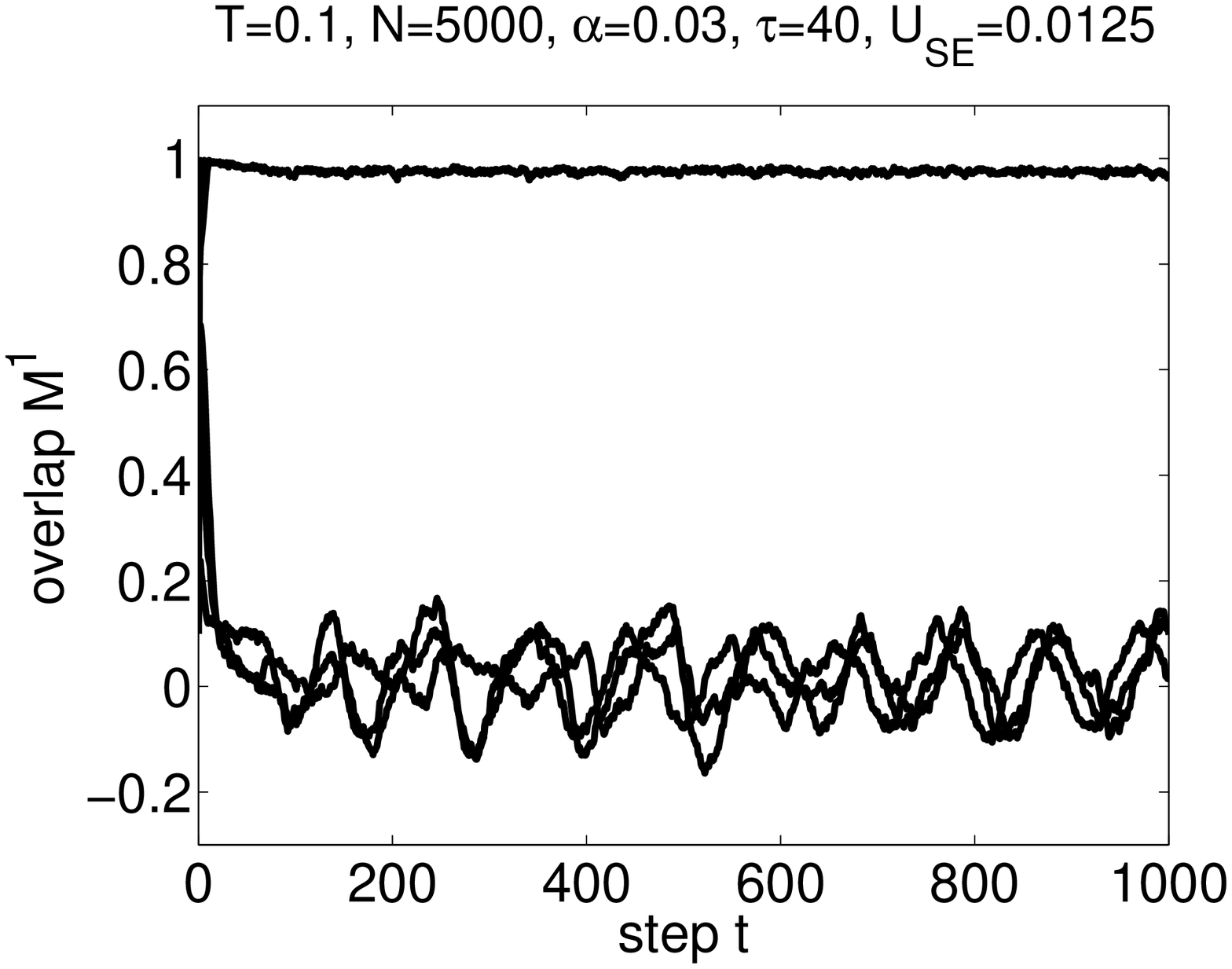}
  \end{tabular}
  \caption{The dependence of overlap $M^1(t)$ on step $t$. The initial overlap $M_0^1$ takes a value in $0.1$, $0.2$, \ldots, $1.0$. (a) The case without synaptic depression. (b) The conditions of case (a) over a long time. (c) The case with synaptic depression, where $\tau=40$ and $U_{SE}=0.0125$. (d) The conditions of case (b) over a long time.}
  \label{fig:overlaps}
 \end{figure}
 
\subsection{Dependence of overlap on step} 
 In this subsection, we show how synaptic depression affects the dependence of the overlap $M^1(t)$ on step $t$.
 Figure \ref{fig:overlaps} shows the dependence obtained from Monte Carlo simulation with $N=5000$, $T=0.1$, and
 loading ratio $\alpha=0.03$, i.e.,\ $p=150$. 
 Initial states are provided with initial overlap $M_0^1$ by varying the value from $0.1$ to $1.0$ at intervals of $0.1$.

 Figure \ref{fig:overlaps}(a) shows the case without synaptic depression from $t=0$ to $t=20$,
 while Fig.\ \ref{fig:overlaps}(b) shows the same case over a long time span, from $t=0$ to $t=1000$.
 When $M_0^1$ is greater than $0.3$, $M^1(t)$ converges to the value $1$ at large $t$, meaning that memory retrieval succeeds.
 This final state corresponds to a memory state.
 On the other hand, when $M_0^1$ is less than $0.4$, $M^1(t)$ is in equilibrium, such that overlaps $M^1(t)$ do not reach the value $1$.
 This equilibrium is called a spurious state. 
 In contrast, Fig.\ \ref{fig:overlaps}(c) shows the case with synaptic depression, where $\tau=40$ and $U_{SE}=0.0125$. 
 When $M^1_0$ is greater than $0.3$, the network succeeds in memory retrieval; otherwise, when $M_0^1$ is less than $0.4$, the network fails in memory retrieval and falls into a spurious state.
 Figure \ref{fig:overlaps}(d) shows the case with synaptic depression over a long time span, from $t=0$ to $t=1000$.
 Here, the overlaps $M^1(t)$ in the spurious state oscillate, whereas those in the memory state converge.
 
 Thus, from the results shown in Fig. \ref{fig:overlaps}, we can see that synaptic depression does not affect memory states but destabilizes spurious states and induces oscillation.
 
\subsection{Autocorrelation functions and period of network dynamics}
 In order to reveal a property of the oscillations shown in Fig.\ \ref{fig:overlaps}(d), we investigate an autocorrelation function $R(k)$ for the overlaps $M^1(t)$.
 This function is a measure of the dependence of $M^1(t)$ on past data and is defined as 
 \begin{equation}
  R(k) = \frac{1}{(L-k)\hat{\sigma}^2}\sum_{t=1}^{L-k}
  \left[M^\mu(t)-\hat{M}\right]\left[M^\mu(t+k)-\hat{M}\right].
 \end{equation}
 Here, $L$ is the number of steps in overlap $M^1(t)$, and $k$ represents the time lag between data.
 The sample mean of overlap $M^1(t)$, $\hat{M}$, and the sample variance, $\hat{\sigma}^2$, are 
 given by the following:
 \begin{eqnarray}
  \hat{M}&=&\frac{1}{L}\sum_{t=1}^LM^\mu(t),\\
  \hat{\sigma}^2&= &\frac{1}{L}\sum_{t=1}^L\left[\hat{M}-M^\mu(t)\right]^2.
 \end{eqnarray}
 
\begin{figure}[t]
  \begin{tabular}{rcrc}\\
   (a)&
   \includegraphics[height=50mm]{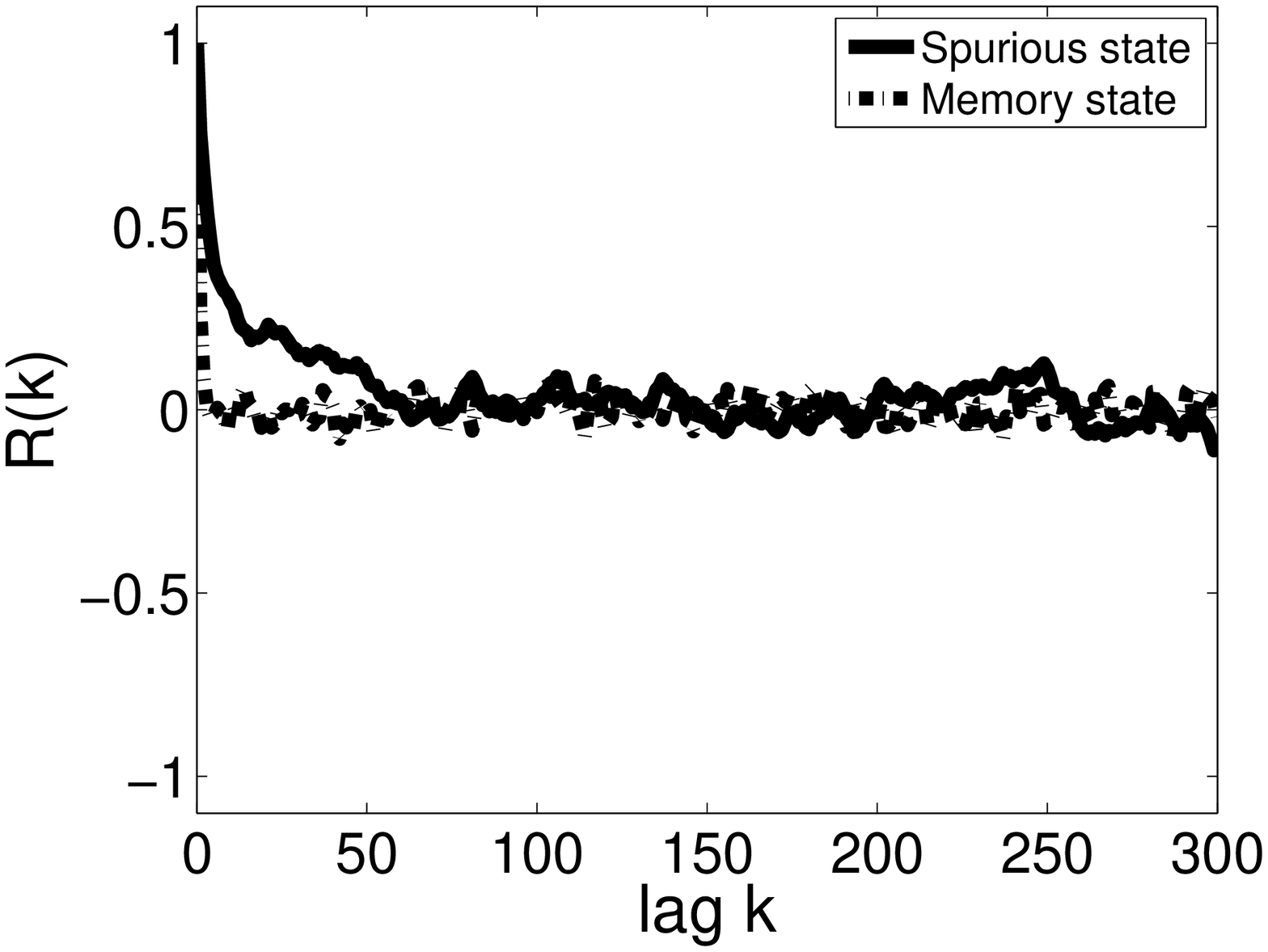}&
   (b)&
   \includegraphics[height=50mm]{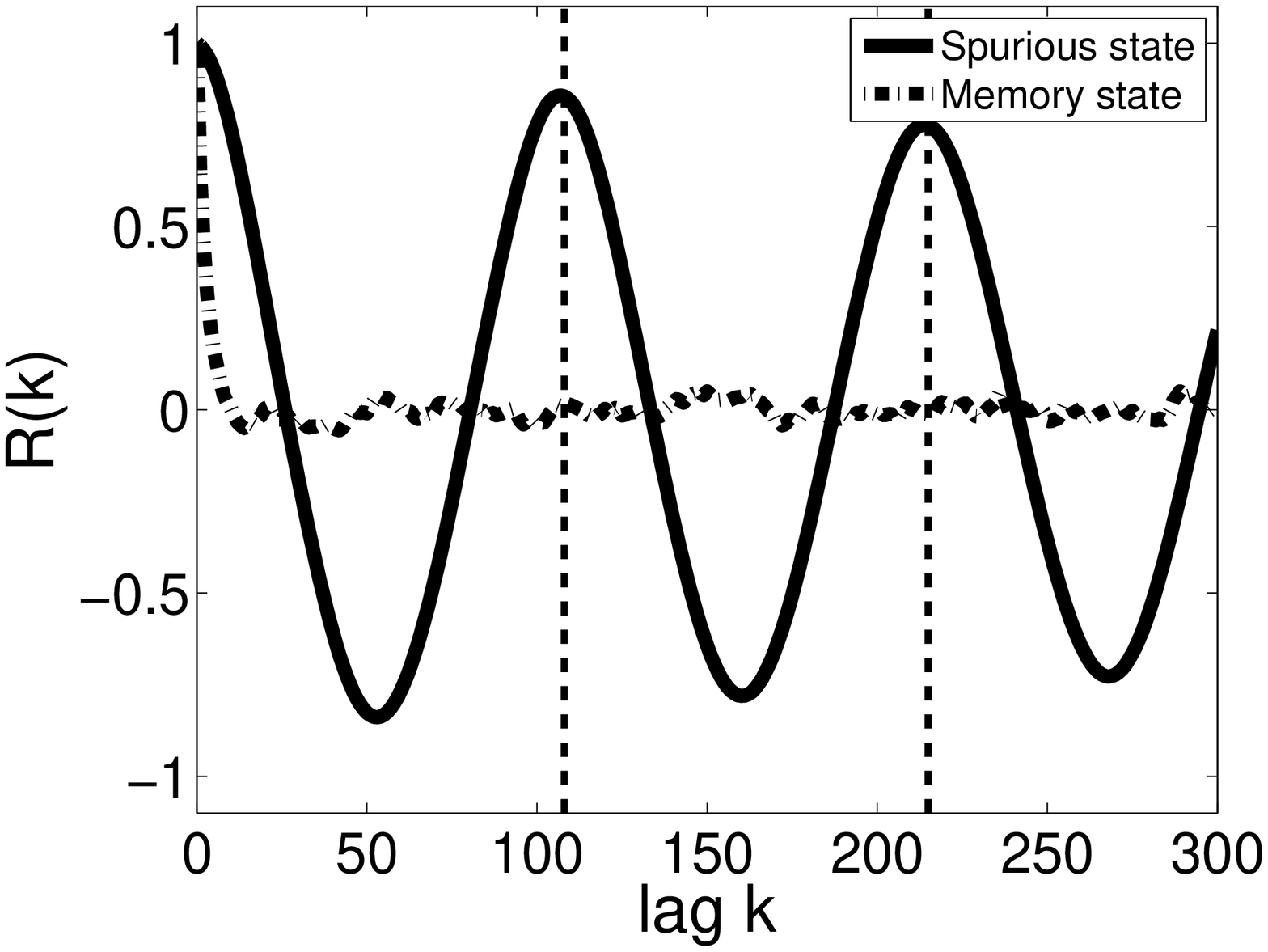}\\
  \end{tabular}
  \caption{Autocorrelation functions, where the dashed lines correspond to memory states and the solid lines correspond to spurious states. (a) The case without synaptic depression. (b) The case with synaptic depression, where $\tau=40$ and $U_{SE}=0.0125$. The orthogonal dotted lines indicate time lags of $k=108$ and $k=215$.}
  \label{fig:overlapac}
 \end{figure}

 Figure \ref{fig:overlapac} shows the dependence of the autocorrelation function $R(k)$ on the lag $k$.
 The dashed lines represent $R(k)$ in a memory state, while the solid lines represent $R(k)$ in a spurious state.
 Note that $M^1_0$ has the values $1.0$ and $0.2$ in the cases of a memory state and a spurious state, respectively.
  
 Figure \ref{fig:overlapac}(a) shows the case without synaptic depression.
 In both the memory state and the spurious state, $R(k)$ rapidly decays to $0$ with increasing $k$. 
 This means that $M^1(t)$ is not affected by past data; rather, its dynamics fluctuate randomly
 because of the probabilistic update given by Eq. (\ref{eq:renew}).
 
 Figure \ref{fig:overlapac}(b) shows the case with synaptic depression, where $\tau=40.0$ and $U_{SE}=0.0125$.
 In the memory state, $R(k)$ decays to $0$ as in the case without synaptic depression.
 On the other hand, in the spurious state, $R(k)$ has high positive amplitude at $k=108$ and $k=215$,
 meaning that $M^1(t)$ oscillates with a period of around $108$ in terms of the time lag.
  
 \begin{figure}[t]
  \begin{center}
   \includegraphics[height=64mm]{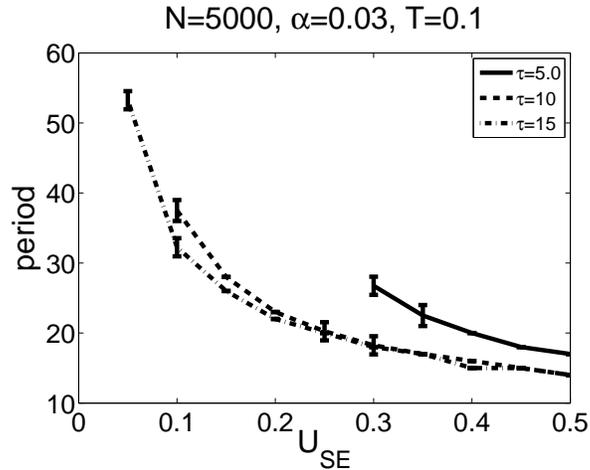}
   \end{center}
  \caption{
  Dependence of period on $U_{SE}$ with $\tau=5.0,10,15$. Each line represents the mean period obtained from five simulations, and each error bar denotes three standard deviations.}
  \label{fig:angle}
 \end{figure}
 Next, we investigate how the period depends on the strength of synaptic depression.
 Figure \ref{fig:angle} shows the dependence of the period on $U_{SE}$.
 Here, $U_{SE}$ varies from $0.05$ to $0.50$ at intervals of $0.05$, while
 $\tau$ is fixed at either $5.0$, $10$, or $15$.  
 Each line in the graph represents the mean period obtained from five simulations, and each error bar indicates three standard deviations.
 We find that as $U_{SE}$ increases, the period decreases and then converges.
 We also see that when $\tau$ is large, it hardly affects the dependence of the period on $U_{SE}$.
 Finally, we find that the period decreases as the strength of the synaptic depression increases.
 
\subsection{Effects on basin of attraction}
\begin{figure}[t]
 \begin{tabular}{rcrc}
  (a)&
  \includegraphics[height=54mm]{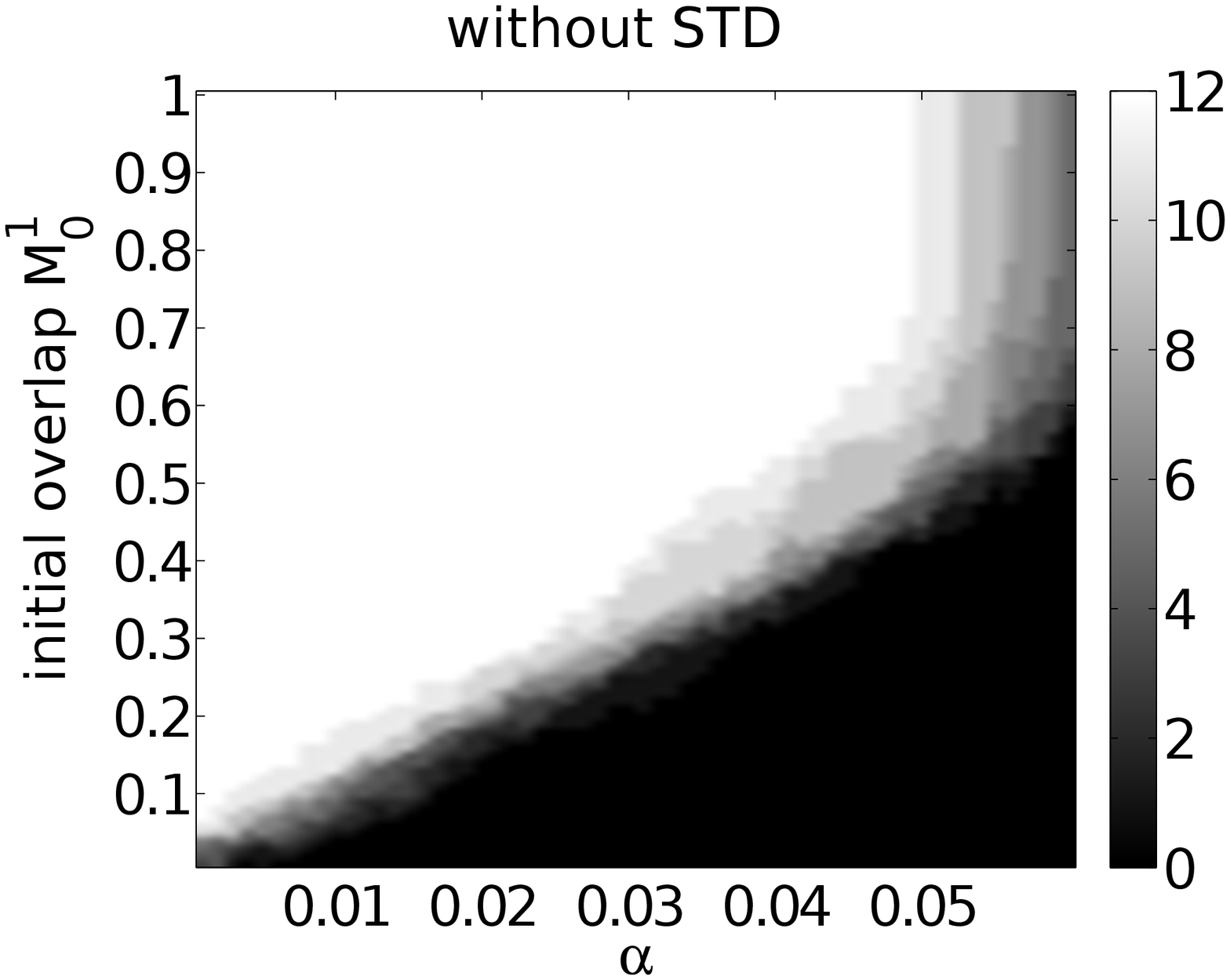}  &
  (b)&
  \includegraphics[height=54mm]{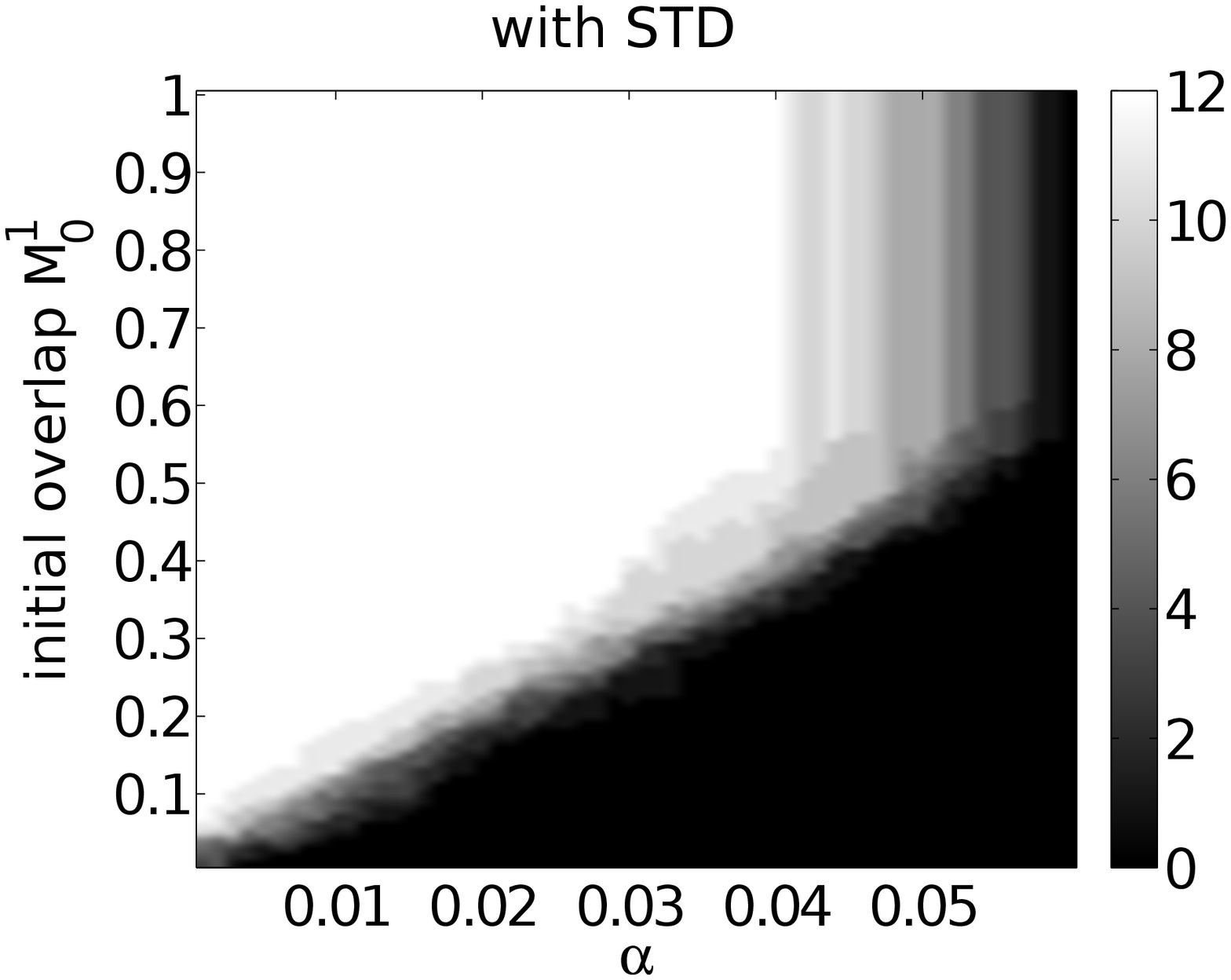}
 \end{tabular}
  \caption{Basins of attraction obtained over 12 simulations with $T=0.1$ and $N=5000$. Each cell represents the number of successful memory retrievals. (a) The case without synaptic depression. (b) The case with synaptic depression, where $\tau=40$ and $U_{SE}=0.0125$.}
  \label{fig:attr}
 \end{figure}

 It turns out that whether memory retrieval succeeds or not depends on the value of the initial overlap $M^1_0$ from Figs.\ \ref{fig:overlaps}(a) and \ref{fig:overlaps}(c).
 There exists a minimum value of the initial overlap $M^1_0$, such that the network succeeds in memory retrieval, and this minimum value is called the critical overlap $M_C$.
 The region where $M^1_0>M_C$ is called a basin of attraction, since the network converges on the memory pattern.
 In this subsection, we investigate the effects of synaptic depression on the basin of attraction.
 Figure \ref{fig:attr} shows basins of attraction obtained from 12 simulations. 
 Each cell represents the number of successful memory retrievals.
 In each trial run, $\alpha$ varies from $0.001$ to $0.06$ at intervals of $0.001$, and $M^1_0$ varies from $0.01$ to $1.0$ at intervals of $0.01$.
 We define the condition that $M^1(t)$ is greater than or equal to $0.8$ at $t=50$ as a successful memory retrieval.
 The boundary between the white and black regions corresponds to the critical overlap $M_C$. 

 Figure \ref{fig:attr}(a) shows the case without synaptic depression.
 It turns out that the value of the critical overlap $M_C$ increases with $\alpha$.
 Furthermore, the value of $M_C$ drastically increases to $1$ at around $\alpha=0.06$.
 This value of $\alpha$ is called the memory storage capacity\cite{Torres2002,Matsumoto2007,Otsubo2011}.
 When $\alpha$ is larger than the memory storage capacity, the network cannot retrieve memory patterns even though it starts from a memory pattern.
 Figure \ref{fig:attr}(b) shows the case with synaptic depression, where $\tau=40$ and $U_{SE}=0.0125$.
 As in the previous case, we see that the value of the critical overlap $M_C$ increases with $\alpha$,
 but in this case $M_C$ drastically increases to $1$ at around $\alpha=0.04$.
 This decrease in memory storage capacity at finite temperature as a result of synaptic depression was revealed previously\cite{Otsubo2011}.
 Note that when $\alpha$ is less than $0.04$, the dependence of $M_C$ on $\alpha$ shows little difference between the cases with and without synaptic depression, as seen by comparing Figs.\ \ref{fig:attr}(a) and \ref{fig:attr}(b).
 This means that when $\alpha$ is less than $0.04$, synaptic depression hardly affects the basin of attraction.
  
  \subsection{Trajectories of network dynamics}
 We find that the overlap $M^1(t)$ in a spurious state oscillates periodically because of synaptic depression, as shown in Fig.\ \ref{fig:overlapac}.
 In order to investigate this oscillating network behavior, we next carry out principal component analysis (PCA) on the network state $\boldsymbol{s}(t)$.
 
 \begin{figure}[t]
  \begin{tabular}{rcrc}
   (a)
   &
   \includegraphics[height=54mm]{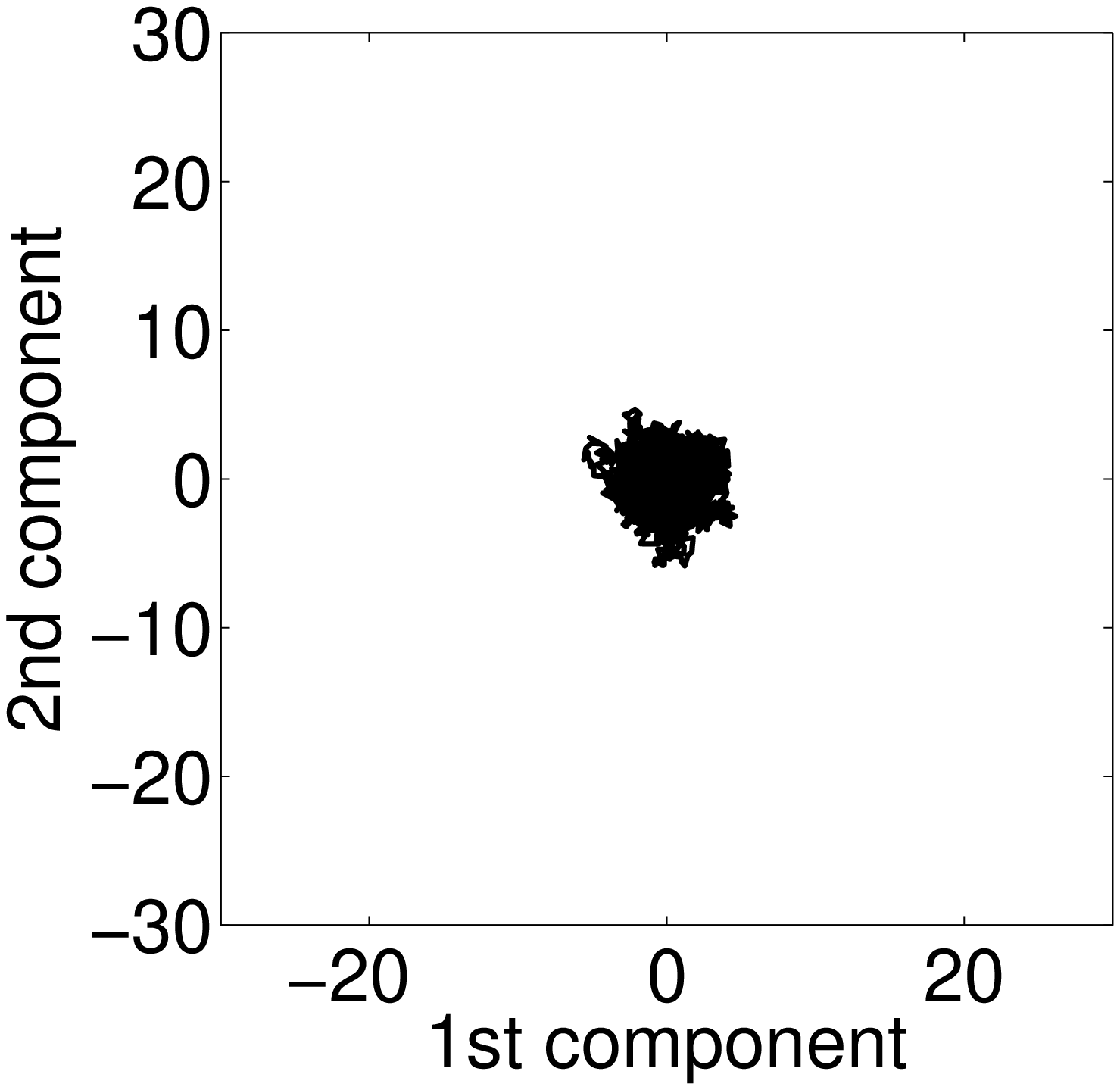}
   &
   (b)
   &
   \includegraphics[height=54mm]{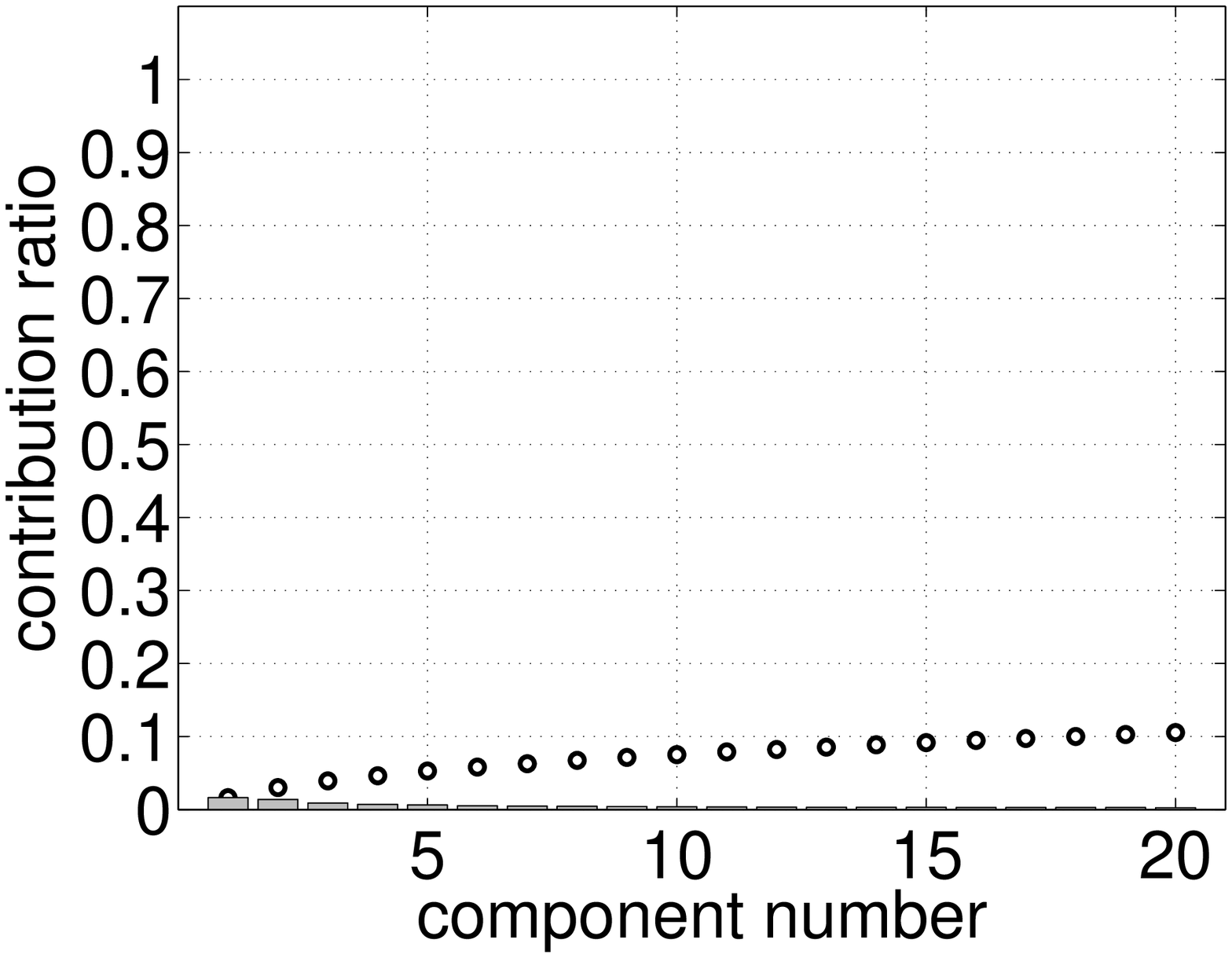} \\
   (c)
   &
   \includegraphics[height=50mm]{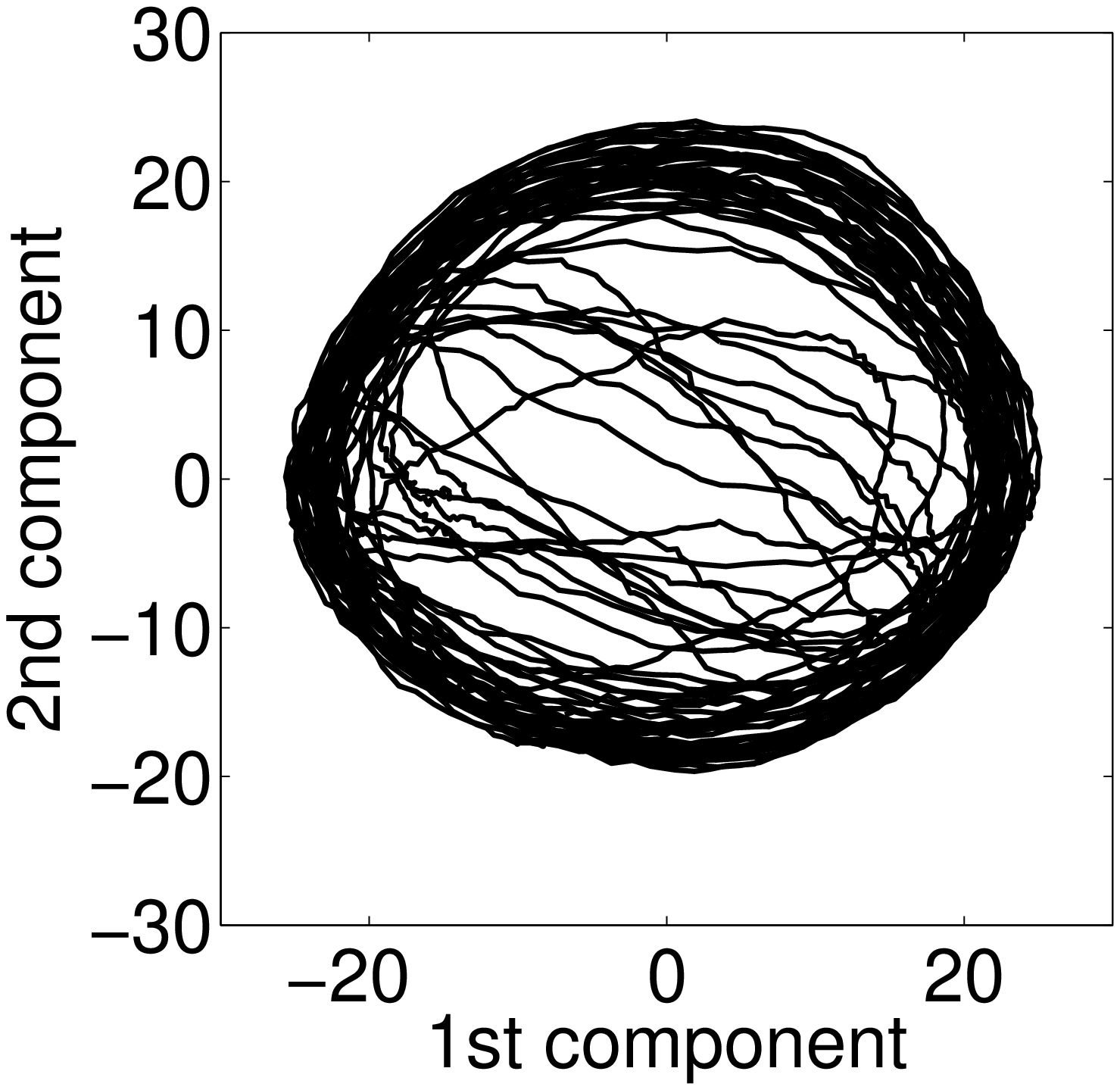}
   &
   (d)
   &
   \includegraphics[height=50mm]{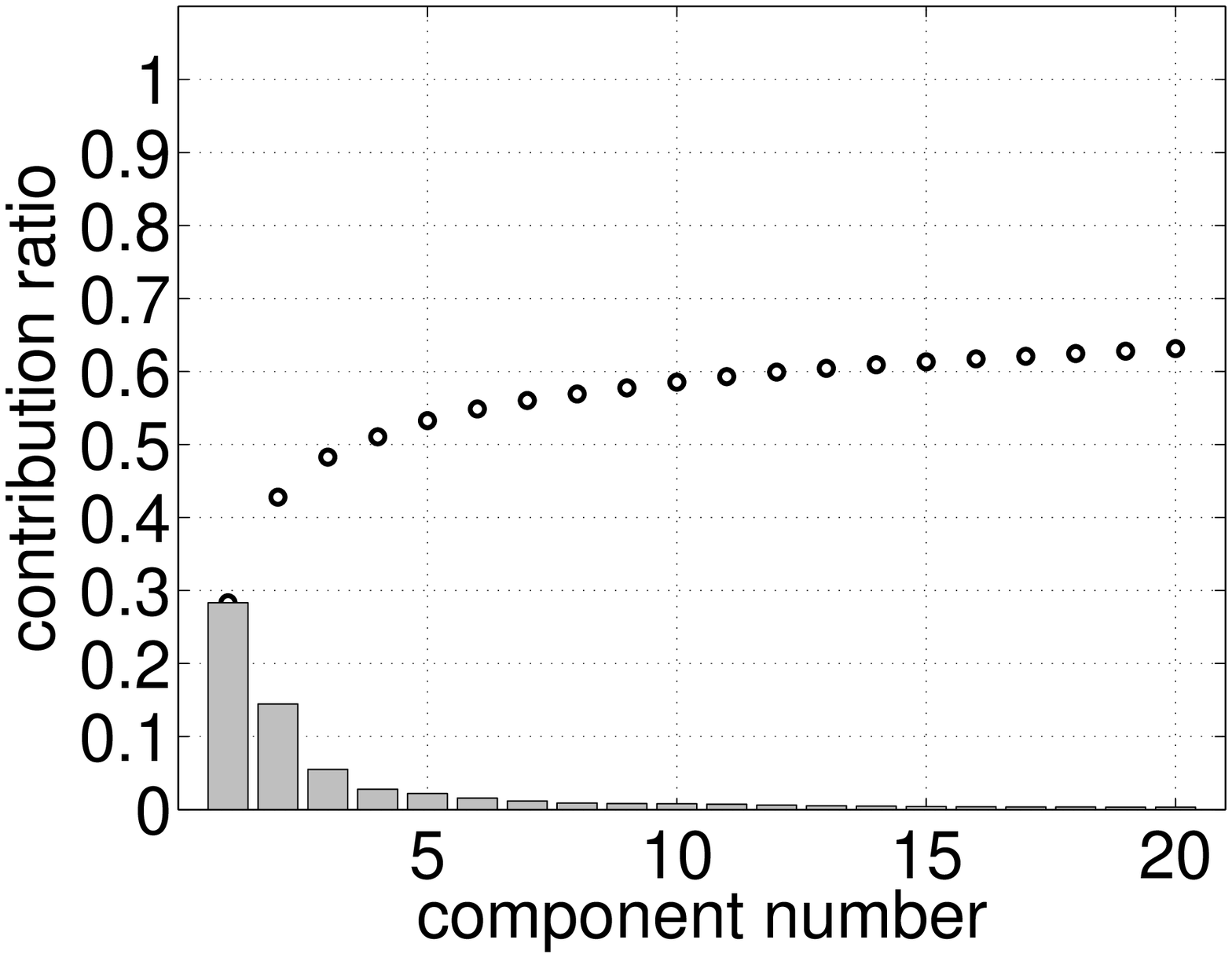}
  \end{tabular}
  \caption{Network trajectories and contribution ratios obtained from PCA, 
  for the cases without (a, b) and with (c, d) synaptic depression, where $\tau=40$ and $U_{SE}=0.0125$ for the latter case of a spurious state. 
  (a, c) Trajectory on a space spanned by the first and second principal components. 
  (b, d) Contribution ratios and cumulative contribution up to the 20th principal component.
  Gray bars and open circles denote median value of contribution ratio and cumulative contribution, respectively.}
  \label{fig:pca}
 \end{figure}
 
 Figure \ref{fig:pca} shows the results obtained from PCA. Specifically, the figure shows the trajectories expressed by the first and second principal components, and the contribution ratios,
 with the parameters $N=5000$, $\alpha=0.03$, $T=0.1$, and $M_0^1=0.2$.

  Figure \ref{fig:pca}(a) shows the case without synaptic depression.
  From the figure, we find that the network fluctuates randomly on the plane spanned by the first and second principal components.
  Figure \ref{fig:pca}(b) shows the contribution ratios and the cumulative contribution up to the 20th principal component.
  All the contribution ratios are lower than $0.05$.

  Figure \ref{fig:pca}(c) shows the case with synaptic depression, where $\tau=40.0$ and $U_{SE}=0.0125$,
 the same parameters used for Fig.\ \ref{fig:overlaps}(d).
  The network trajectory has circular motion on the plane spanned by the first and second principal components.
  Figure \ref{fig:pca}(d) shows the contribution ratios and cumulative contribution up to the 20th principal component for the case with synaptic depression.
  The contribution ratios of both the first and second principal components are particularly high, with a sum of around $0.42$ 

 \begin{figure}[b]
 \begin{tabular}{rcrc}
  (a)&
  \includegraphics[height=50mm]{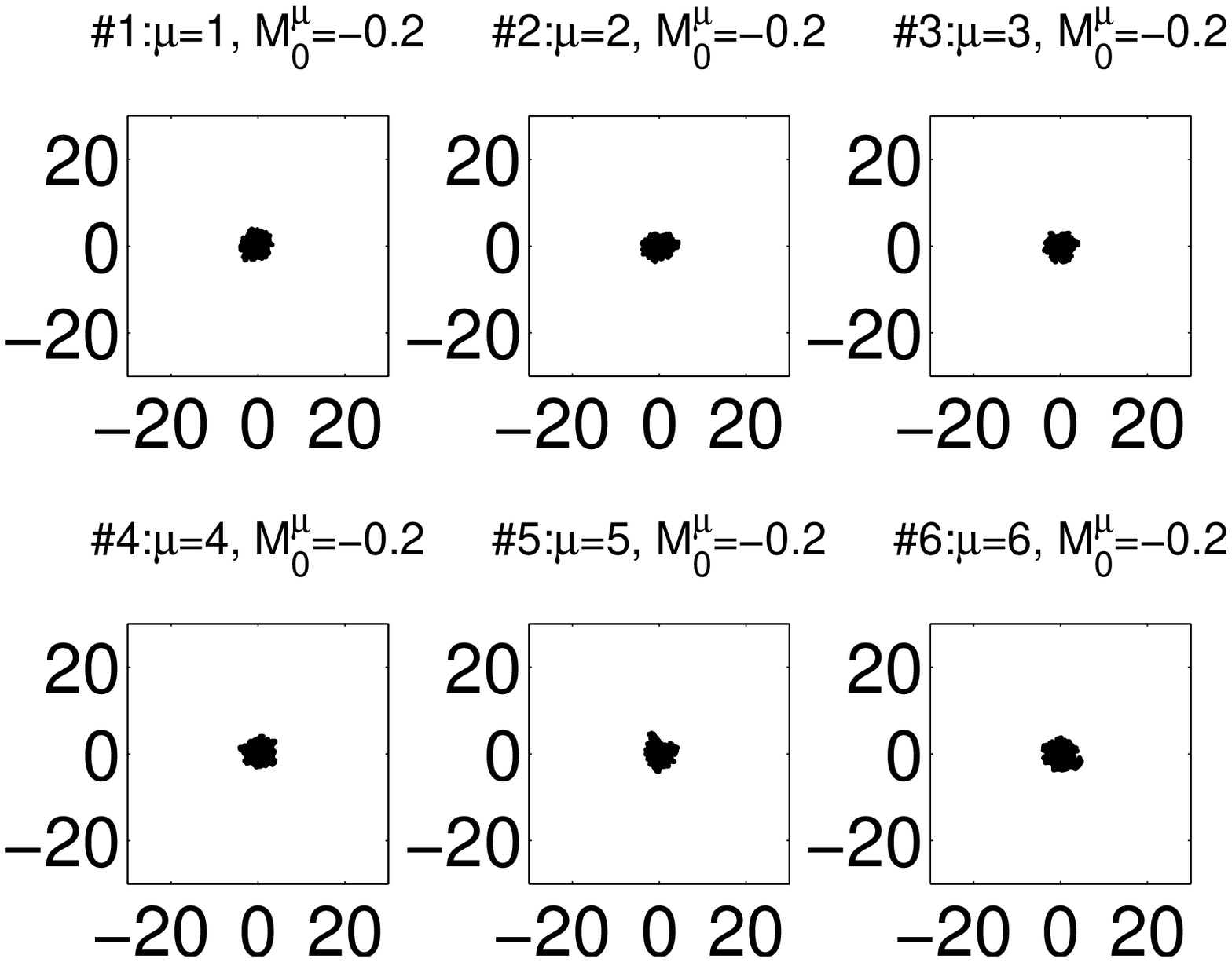}
  (b)&
   \includegraphics[height=50mm]{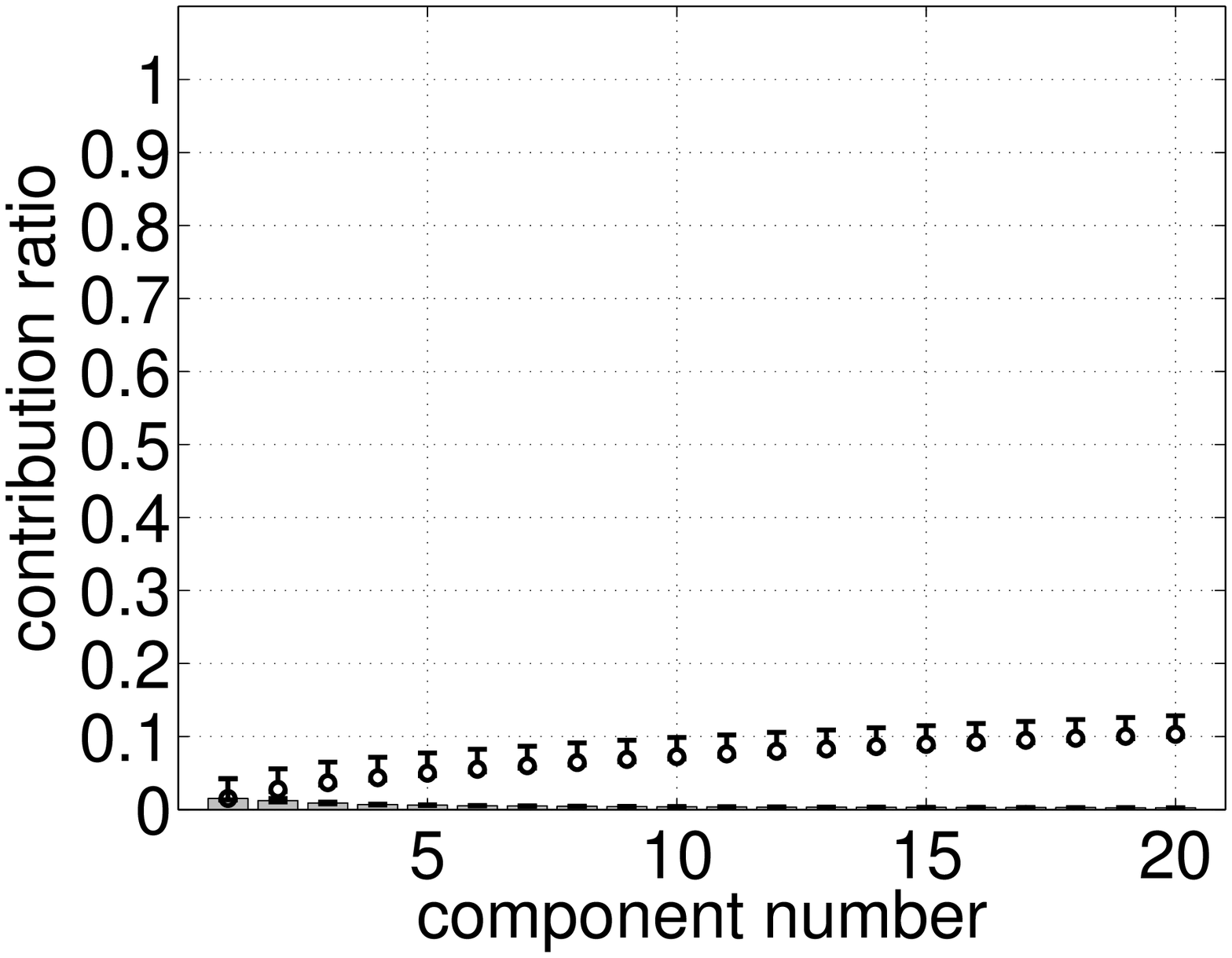}\\
  (c)&
  \includegraphics[height=50mm]{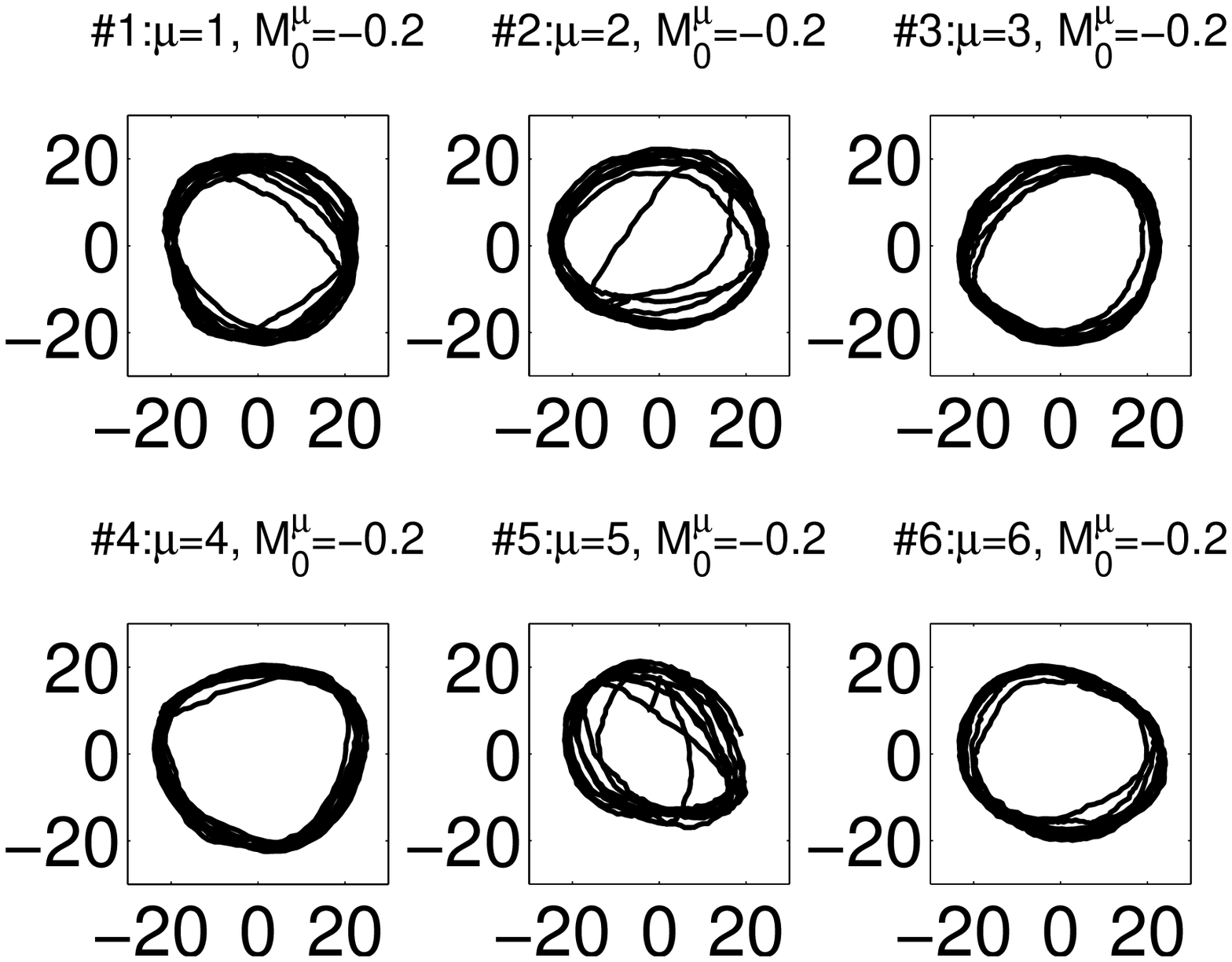}
  (d)&
   \includegraphics[height=50mm]{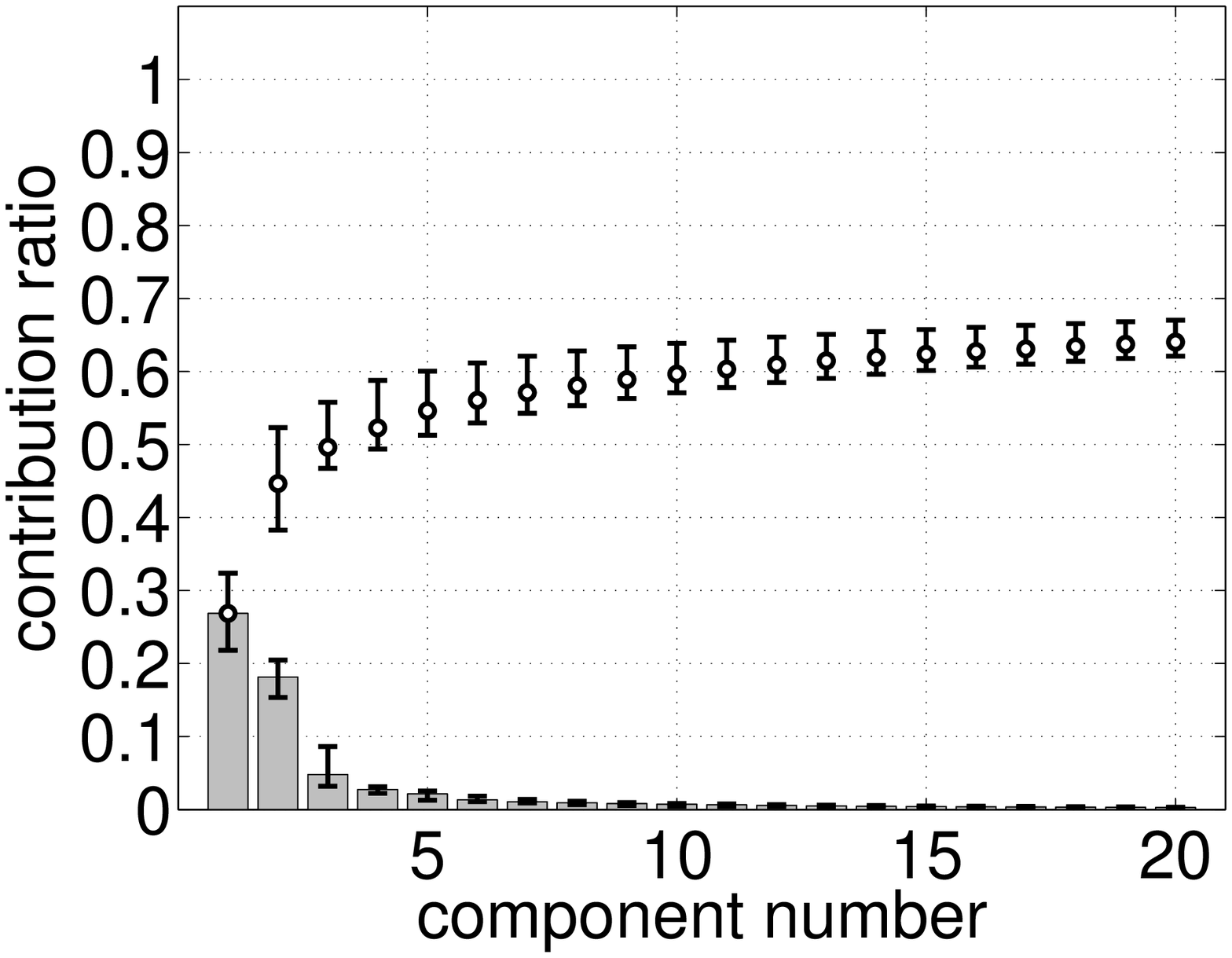}
 \end{tabular}
  \caption{
  Results obtained from PCA on 750 data with different initial conditions,
  for the case without (a, b) and with (c, d) synaptic depression, where $\tau=40.0$, and $U_{SE}=0.0125$ for the latter case.
  (a, c) Typical trajectories on a space spanned by the first and the second principal components.
  (b, d) Contribution ratio and cumulative contribution up to the 20th principal component.
  Gray bars and open circles denote median value of contribution ratio and cumulative contribution, respectively.
  Error bars represent 5th-95th percentile range.
  }
  \label{fig:contratioerrbar}
 \end{figure}
 
 Next, we change the value of initial pattern, $\mu$, and initial overlap $M^\mu_0$, and then carry out 750 simulations,
 i.e., $\mu$ varying from $1$ to $150$ at intervals of $1$, and $M^\mu_0$ varying from $-0.2$  to $0.2$ at intervals of $0.1$.
 We perform $15000$-step simulations, and then count out first $10000$ steps as initial relaxation.
 Furthermore, in the case without synaptic depression, the network state in two simulations converge on memory patterns.
 Thus, we count out these data.
 We show the result of PCA on data set obtained from these simulations in Figs. \ref{fig:contratioerrbar}.
 Figures \ref{fig:contratioerrbar}(a) and \ref{fig:contratioerrbar}(b) show the case without synaptic depression.
 Figure \ref{fig:contratioerrbar}(a) shows  network trajectories on a plane spanned by the first and the second principal components.
 We show the $6$ typical results from $748$ trials.
 Bars and open circles in the Fig. \ref{fig:contratioerrbar} (b) represent the median value of contribution ratio, and cumulative contribution, respectively.
 Error bars represent 5th-95th percentile range.
 We find that the networks fluctuate randomly on the high-dimensional subspace, because almost all contribution ratios are less than $0.05$.
 Figures \ref{fig:contratioerrbar}(c) and \ref{fig:contratioerrbar}(d) show the case with synaptic depression, where $\tau=40.0$, and $U_{SE}=0.0125$.
 Figure \ref{fig:contratioerrbar}(c) shows the network trajectories on plane consisting of the first and the second principal components.
 We show the $6$ typical results from $750$ trials. 
 Bars and open circles in the Fig. \ref{fig:contratioerrbar}(d) represent the median value of contribution ratio, and cumulative contribution, respectively. 
 Error bars represent 5th-95th percentile range. 
 We find that the network in spurious states predominantly oscillate on the plane spanned by the first and the second principal components, 
 because contribution ratio of the first and the second principal components are higher than that of other principal component.
From the result shown in Fig. \ref{fig:contratioerrbar},
we can see that network in spurious states with synaptic depression oscillates circularly on low-dimensional plane,
whereas network in spurious states without synaptic depression fluctuates randomly.

\begin{figure}[b]
	\begin{tabular}{rcrc}
		(a)&
		\includegraphics[width=64mm]{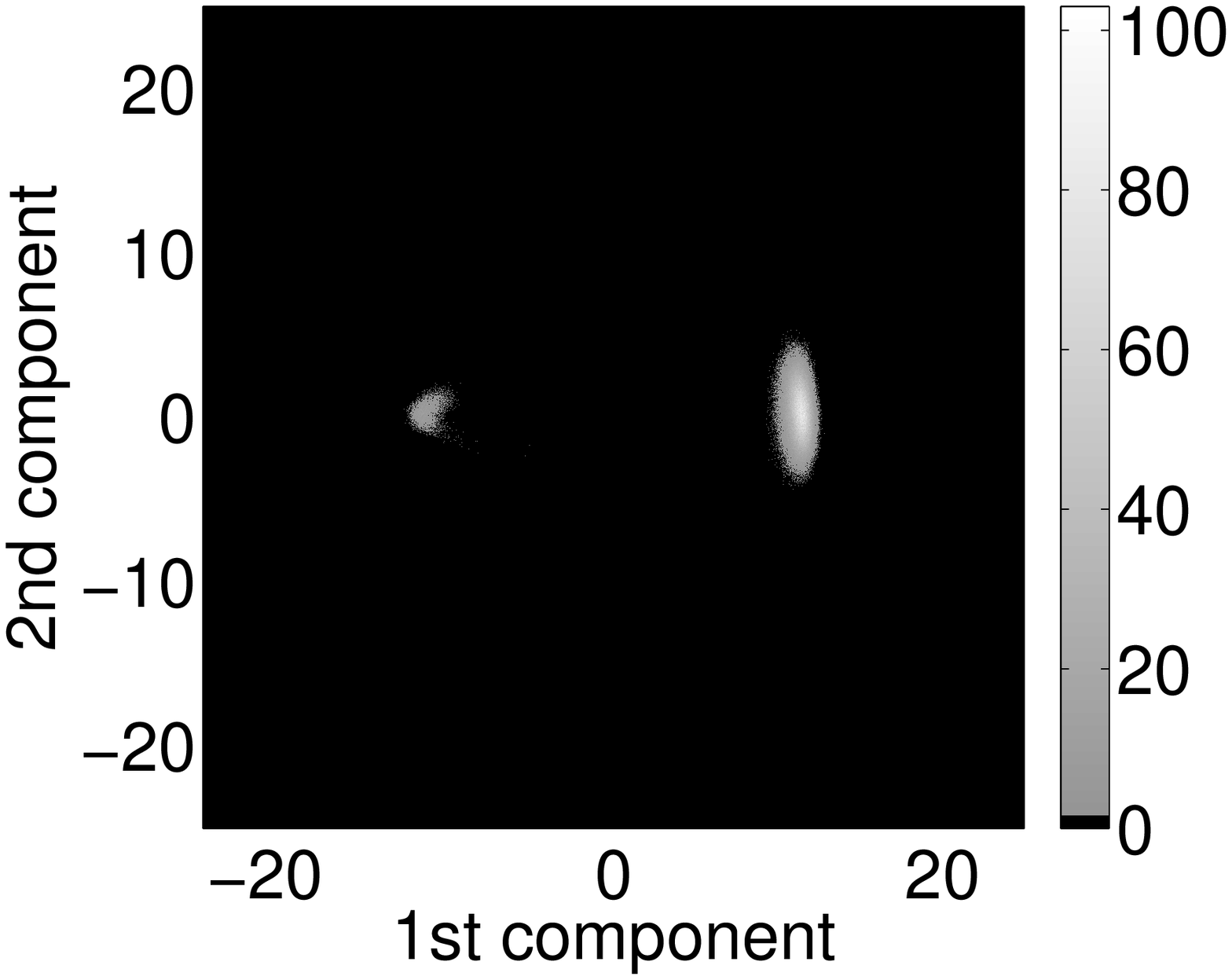}&
		(b)&
		\includegraphics[width=64mm]{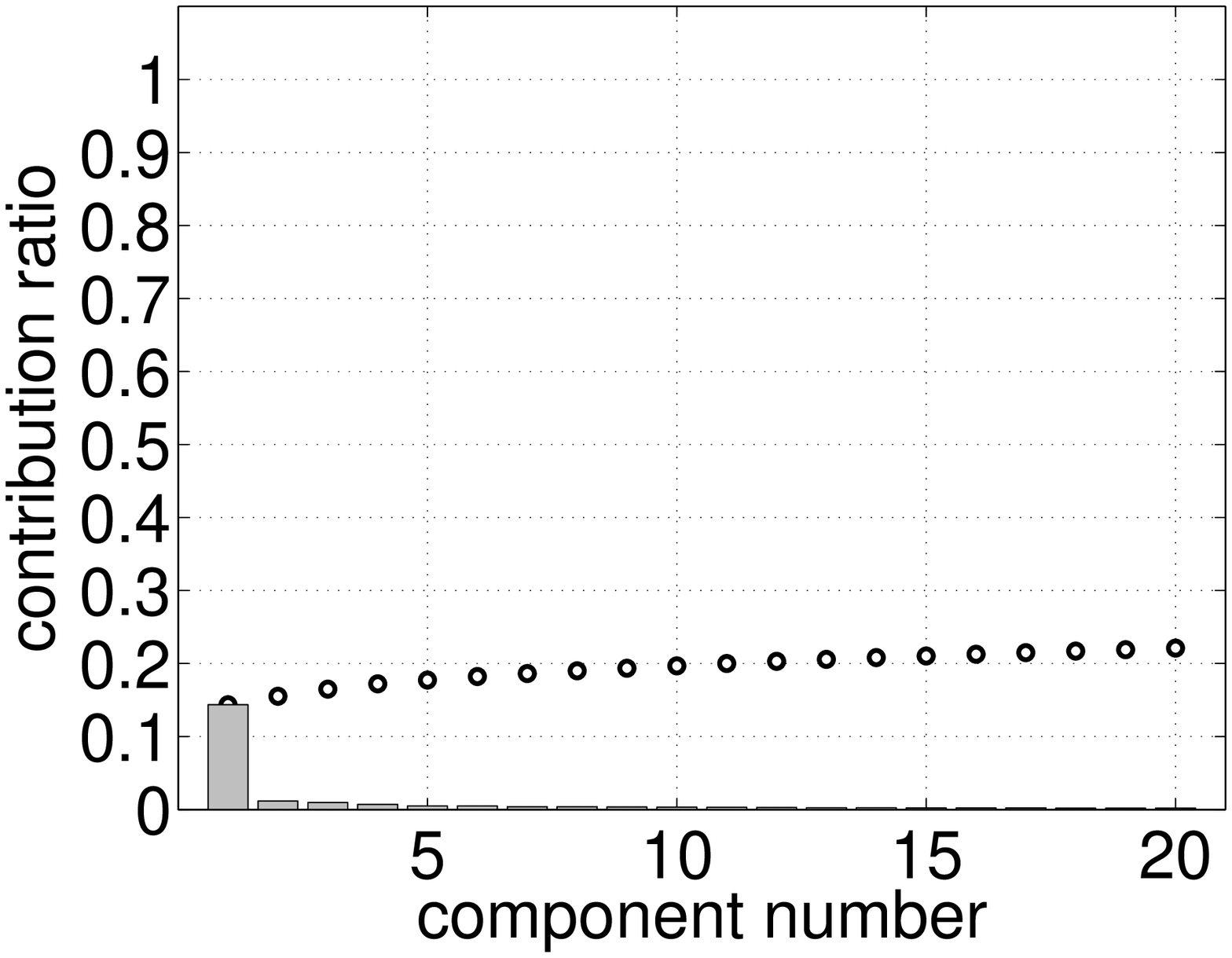}\\
		(c)&
		\includegraphics[width=64mm]{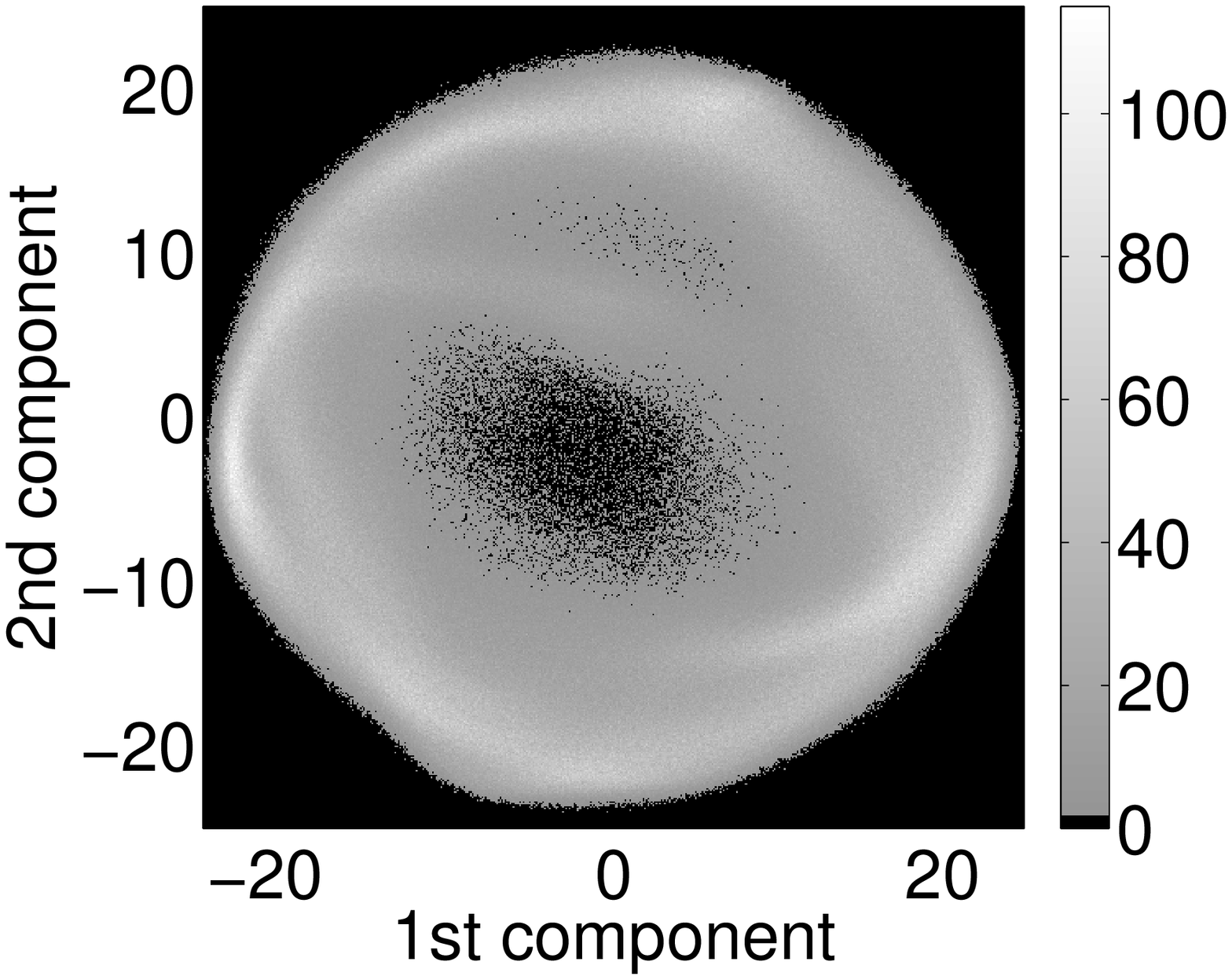}&
		(d)&
		\includegraphics[width=64mm]{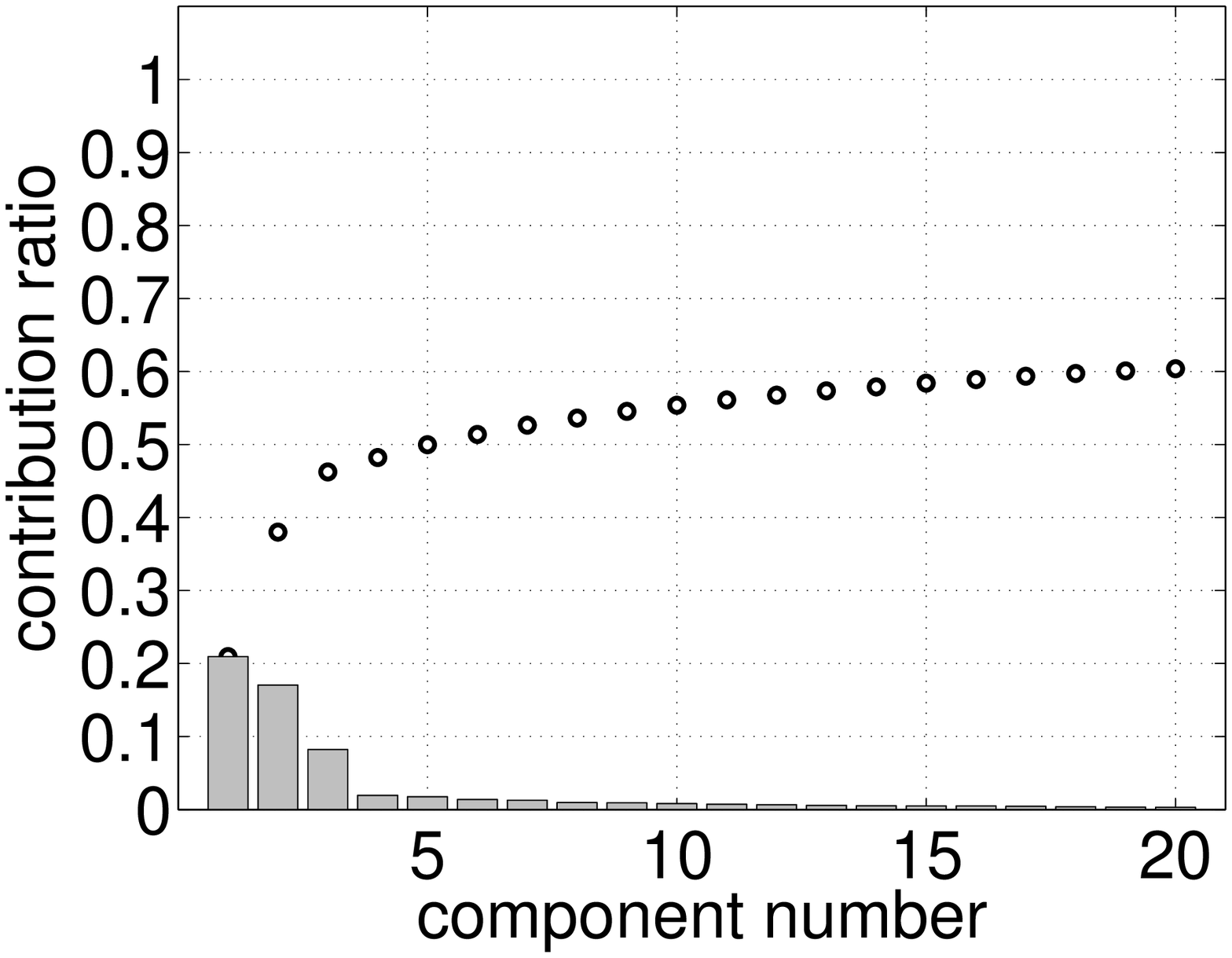}
	\end{tabular}
	\caption{
	Result of PCA on combined data set consisting of 750 different initial conditions,
	for the case without (a, b) and with (c, d) synaptic depression, where $\tau=40.0$, and $U_{SE}=0.0125$ for the latter case.
	(a, c) Histogram of plane spanned by the first and the second principal components.
	Brightness represents the number of spurious states.
	(b, d) Gray bars and open circles correspond to contribution ratio and cumulative contribution, respectively.
	}
	\label{fig:allpca}
\end{figure}

Next, in order to investigate relationship among trajectories, we combine all network state data, $\boldsymbol{s}(t)$, obtained from 750 different initial condition, and then carry out PCA on the data.
Figures \ref{fig:allpca}(a) and \ref{fig:allpca}(b) show the result obtained from PCA in the case without synaptic depression.
Figure \ref{fig:allpca}(a) shows distribution of spurious states on plane spanned by the first and the second principal components as histogram.
The distribution is shown as gray scale image, with black and white corresponding to the smallest and the largest number of spurious states, respectively.
Figure \ref{fig:allpca}(b) shows contribution ratio and cumulative contribution.
Bar denotes contribution ratio, and open circle denotes cumulative contribution.
We can see that contribution ratio of the first principal component in this combined data case is higher than separate data case.
From these results, spurious states in the case without synaptic depression are separated on the plane spanned by the first and the second principal components.
This result is similar to behavior of spin-glass state data of Sherrington-Kirkpatrick model obtained from PCA analysis\cite{Inoue2005}.
Figures \ref{fig:allpca}(c) and \ref{fig:allpca}(d) show the result obtained from PCA in the case with synaptic depression, where $\tau=40.0$, and $U_{SE}=0.0125$.
Figure \ref{fig:allpca}(c) shows distribution of spurious states on plane spanned by the first and the second principal components as histogram.
The distribution is shown as gray scale image, with black and white corresponding to the smallest and the largest number of spurious states, respectively.
Figure \ref{fig:allpca}(d) shows contribution ratio and cumulative contribution.
Bar denotes contribution ratio, and open circle denotes cumulative contribution.
Contribution ratio up to the third principal component is higher than other principal components.
We see that network in spurious states oscillates in low-dimensional subspace, i.e. around three dimension, in the case with synaptic depression.

\subsection{Eigenvectors of synaptic weight matrix}
From the above, we see that network in spurious states fluctuates in high-dimensional subspace in the case without synaptic depression,
whereas network oscillates in low-dimensional subspace in the case with synaptic depression.
In the former case, dynamics of memory recalling in the associative memory model was discussed qualitatively by using eigenvectors of synaptic weight matrix.
Kindo and Kakeya claimed that linear transformation of synaptic weight matrix is important to analyze dynamical properties\cite{Kindo1998}.
Thus, in this subsection, we investigate relationship between subspace in which oscillation induced by synaptic depression occurs and subspace spanned by eigenvectors of synaptic weight matrix.

We consider synaptic weight matrix $\boldsymbol{J}$ whose $(i,j)$ entry is the static weight $\tilde{J}_{ij}$.
Eigenvector $\boldsymbol{v}_n$ of $\boldsymbol{J}$ corresponds to the $n$-th largest eigenvalue $\lambda_n$.
Here, $\boldsymbol{J}$ has $p$ positive eigenvalues and $(N-p)$ zero eigenvalues.
Considering component of $\boldsymbol{J}$, we see that subspace spanned by eigenvectors corresponding to positive eigenvalues is composed of memory patterns, 
and subspace spanned by eigenvectors corresponding to zero eigenvalues is orthogonal to subspace spanned by memory patterns.

\begin{figure}[b]
	\begin{tabular}{rcrc}
		(a)&
		\includegraphics[width=64mm]{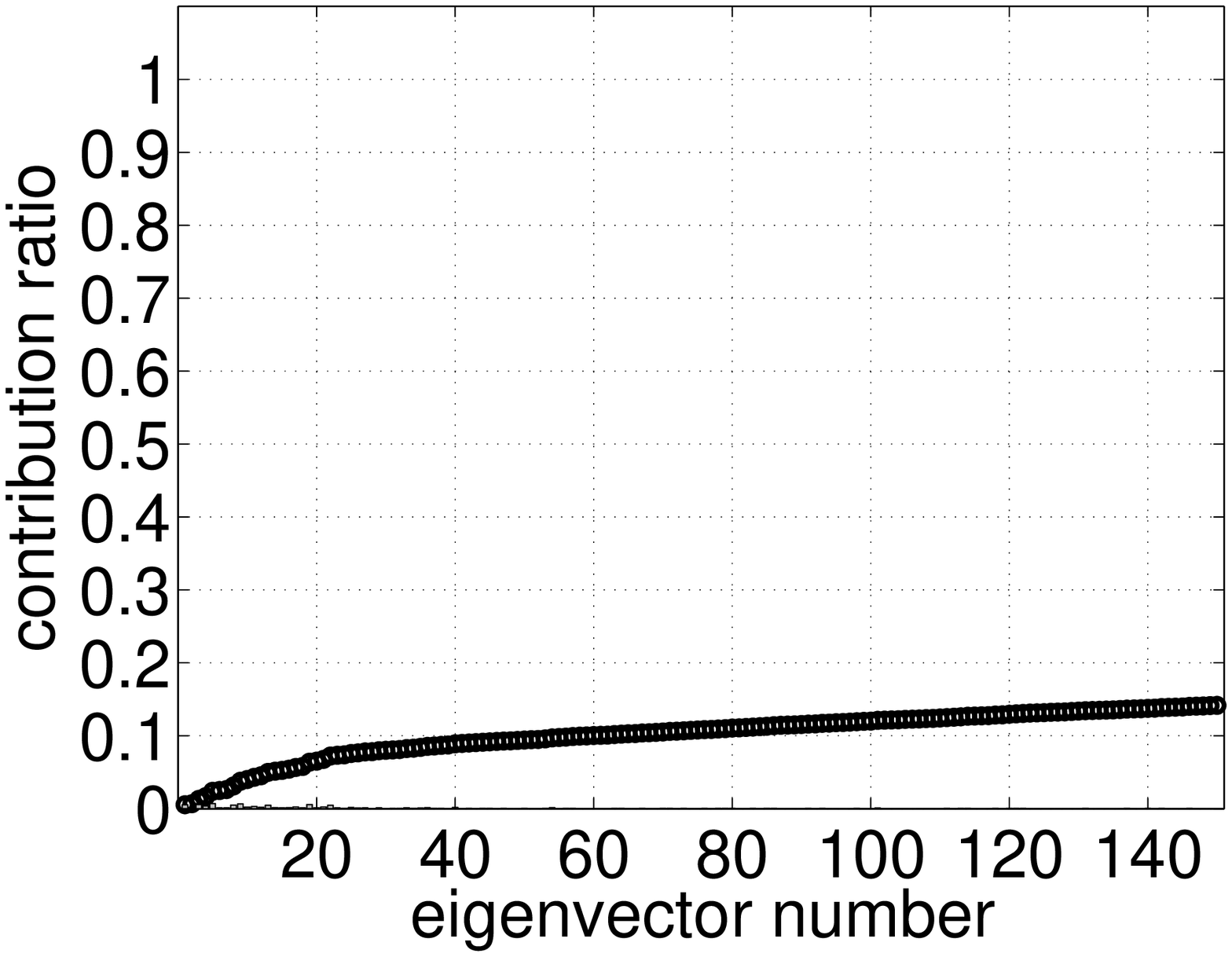}&
		(b)&
		\includegraphics[width=64mm]{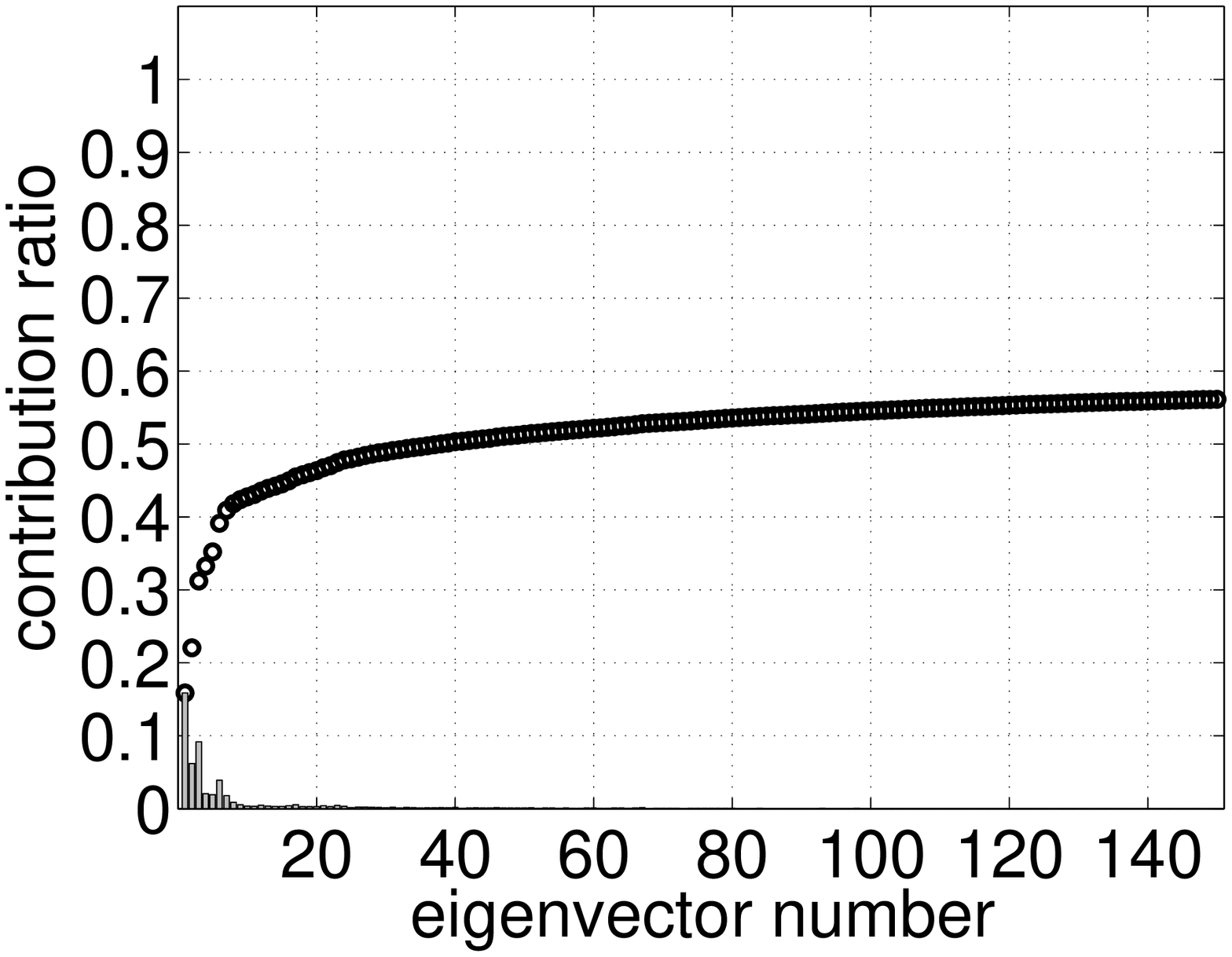}	
	\end{tabular}
	\caption{
	Contribution ratio and cumulative contribution when the network state is expressed by using eigenvector, $\boldsymbol{v}_n$, of static weight matrix, $\boldsymbol{J}$.
	Bars and open circles represent contribution ratio and cumulative contribution of $\boldsymbol{v}_n$ up to $n=150$, respectively.
	(a) The case without synaptic depression.
	(b) The case with synaptic depression, when $\tau=40.0$, and $U_{SE}=0.0125$.
	}
	\label{fig:comparecont}
\end{figure}

Network state $\boldsymbol{s}(t)$ is expressed by using $\{\boldsymbol{v}_n\}$ as
\begin{eqnarray}
	\boldsymbol{s}(t)&=&\sum_{n=1}^{N}a_n(t)\boldsymbol{v}_n,\\
	a_n(t)&=&\boldsymbol{v}_n^\mathrm{T}\boldsymbol{s}(t).
\end{eqnarray}
Contribution ratio, $r_n$, of each $\boldsymbol{v}_n$ is defined as
\begin{equation}
	r_n=\frac{\mathrm{var}[a_n(t)]}{\sum_m\mathrm{var}[a_m(t)]}.
\end{equation}
Figure \ref{fig:comparecont} shows contribution ratio, $r_n$, and cumulative contribution up to $n=150$,
since $\boldsymbol{J}$ has 150 positive eigenvalues and 4850 zero eigenvalues in this study.
Bars and open circles denote contribution ratio and cumulative contribution, respectively.
Figure \ref{fig:comparecont}(a) shows the case without synaptic depression, 
and Fig. \ref{fig:comparecont}(b) shows the case with synaptic depression, where $\tau=40.0$, and $U_{SE}=0.0125$.
In the case without synaptic depression, we see that spurious states mainly exist in subspace which is orthogonal to memory patterns, since cumulative contribution is around $0.14$ at $n=150$.
In the case with synaptic depression, however, we see that spurious states mainly exist in subspace which is spanned by memory patterns, since cumulative contribution is around $0.57$ at $n=150$.
From the results shown in Fig. \ref{fig:comparecont}, we see that oscillation of network induced by synaptic depression mainly occurs on subspace spanned by memory patterns.

  \begin{figure}[b]
  \begin{tabular}{rcrc}\\
	(a)&
    \includegraphics[width=64mm]{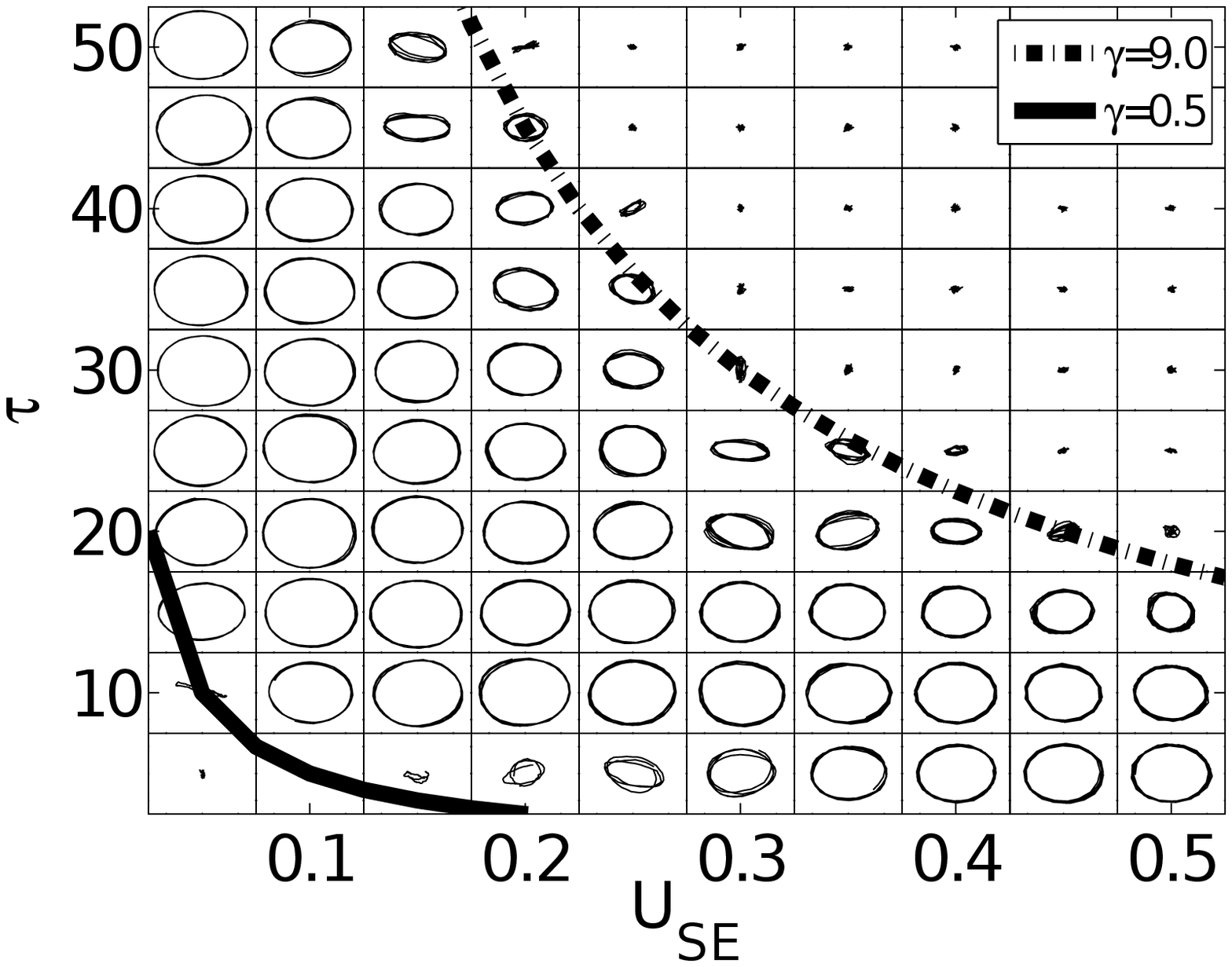} &
   (b)&
   \includegraphics[width=70mm]{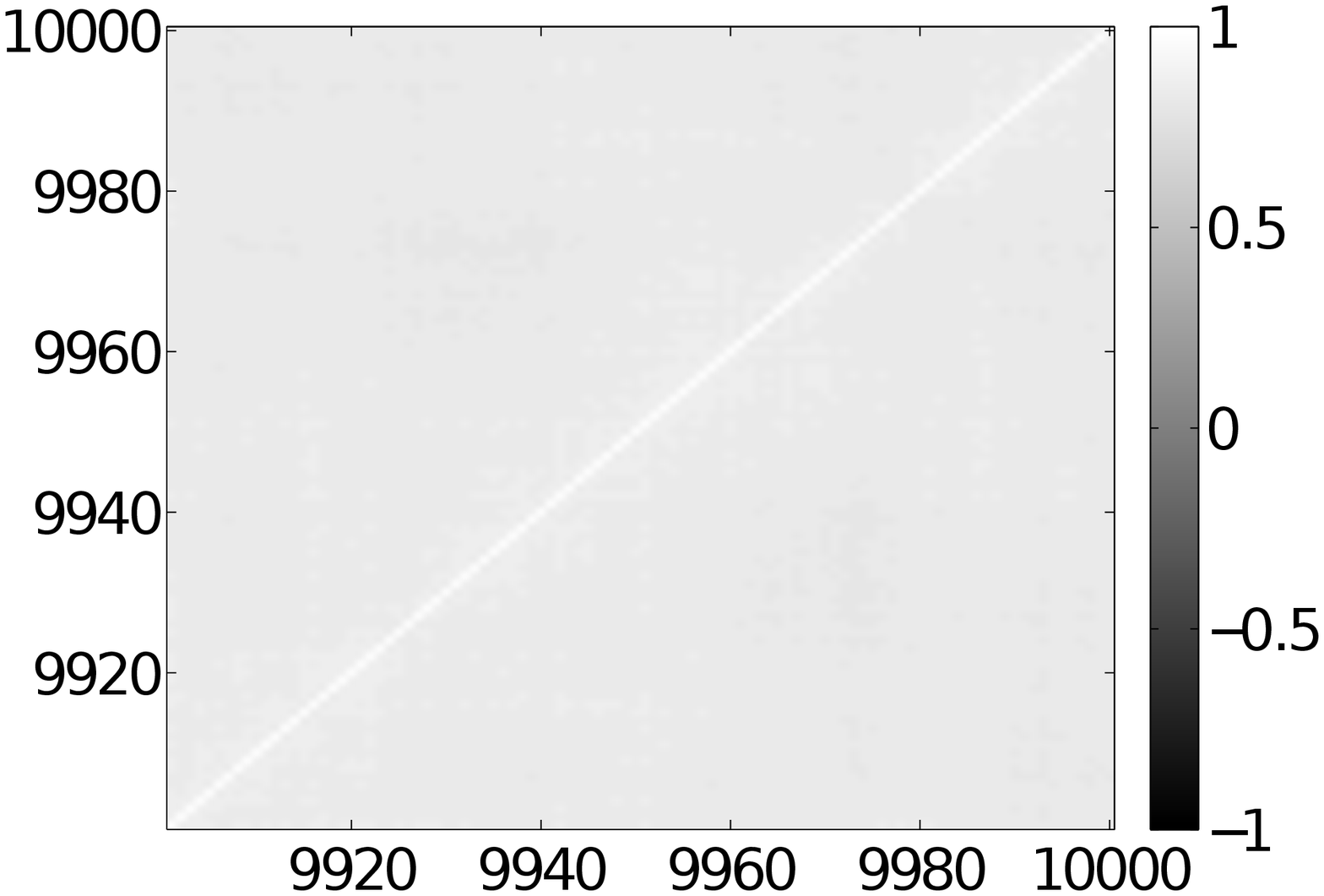} \\
	(c)&
    \includegraphics[width=70mm]{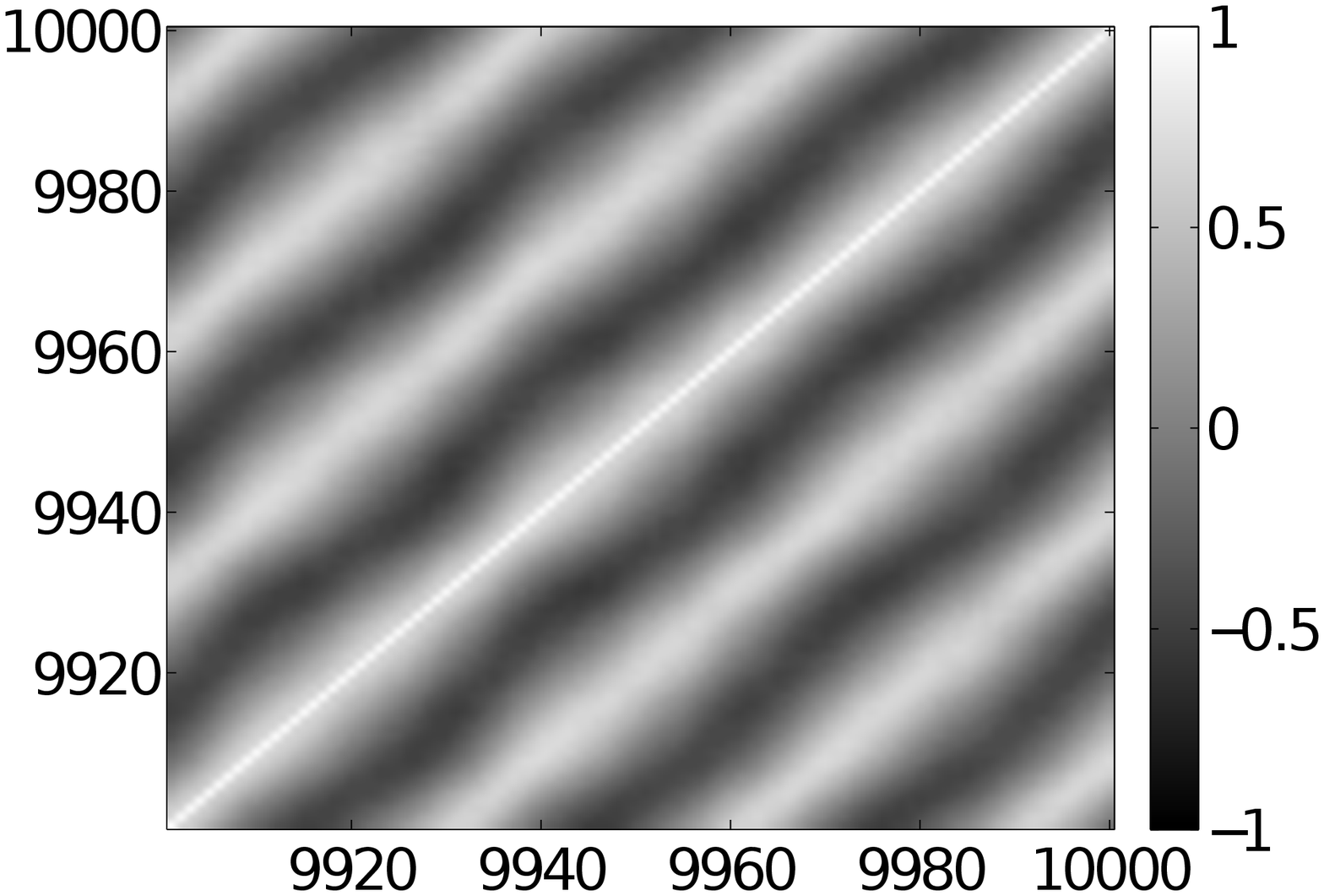} &
   (d)&
    \includegraphics[width=70mm]{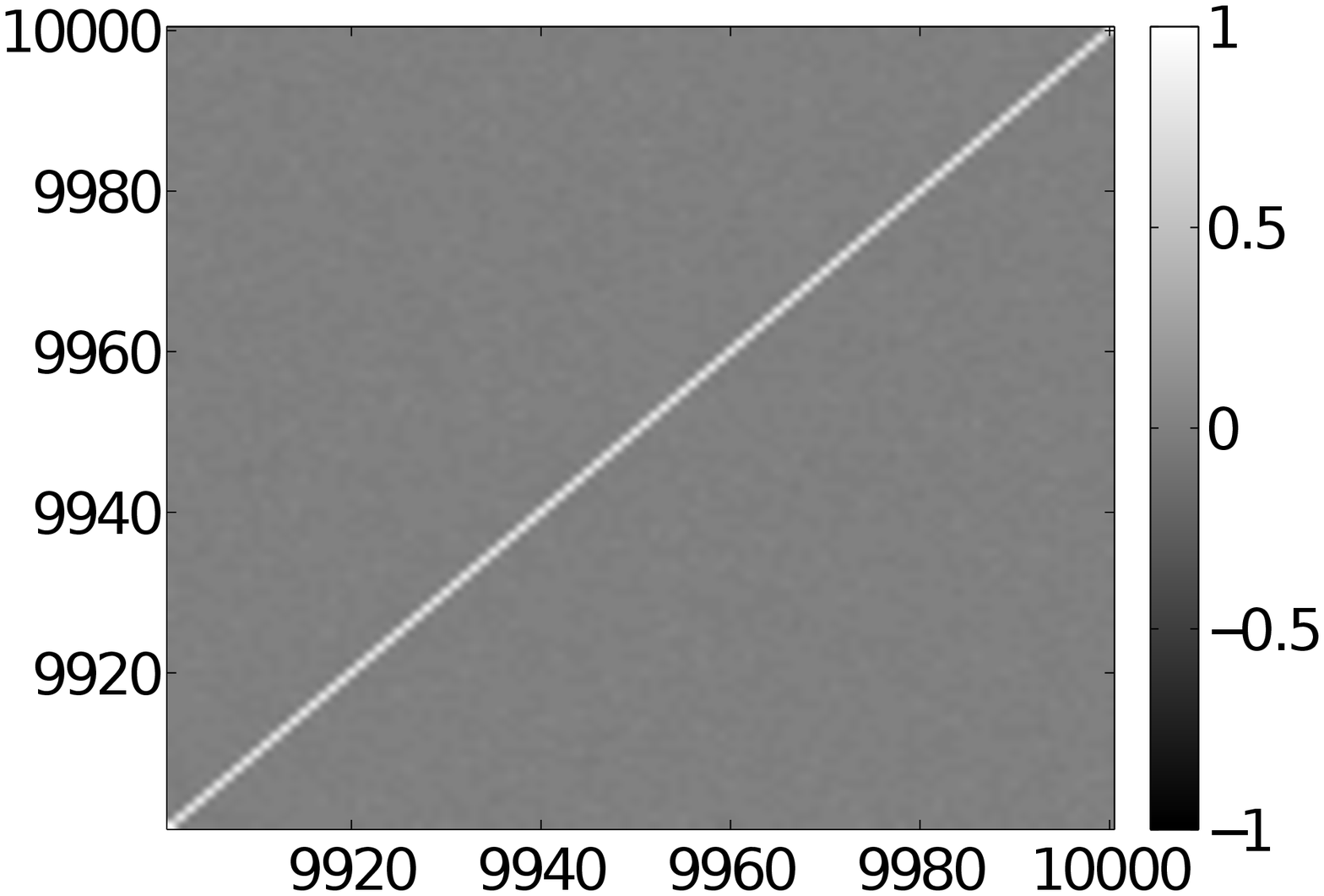}    \\
   \end{tabular}
   \caption{(a) Phase diagram for network dynamics in spurious states. (b-d) Overlaps between network states, for the cases of (b) $\tau=5.0$ and $U_{SE}=0.05$, (c) $\tau=20$ and $U_{SE}=0.1$, and (d) $\tau=50$ and $U_{SE}=0.5$.}
   \label{fig:phdiag}
  \end{figure}
    
  \subsection{Phase diagram}
  In this subsection, we investigate how the network trajectory depends on the strength of synaptic depression.
  Figure \ref{fig:phdiag}(a) shows a phase diagram for this dependence.
  The horizontal axis corresponds to $U_{SE}$, and the vertical axis corresponds to $\tau$.
  The value of $U_{SE}$ varies from $0.05$ to $0.5$ at intervals of $0.05$,
  and $\tau$ varies from $5.0$ to $50$ at intervals of $5.0$.
  Each cell represents the network trajectory on the plane spanned by the first and second principal components.
  The solid and dashed lines correspond to $\gamma=0.5$ and $\gamma=9.0$, respectively,
  where $\gamma$ represents the strength of synaptic depression in equilibrium and is defined as $\gamma = \tau U_{SE}$\cite{Pantic2002,Otsubo2011}.
  The trajectories have circular motion when $\gamma$ is more than $0.5$ and less than or equal to $9.0$.

  We now consider the overlap $C(t_m,t_n)$ between network states $\boldsymbol{s}(t_m)$ and $\boldsymbol{s}(t_n)$, given by:
  \begin{equation}
  C(t_m,t_n) = \frac{1}{N}\sum_{i=1}^N\left(2s_i(t_m)-1\right)\left(2s_i(t_n)-1\right).
  \end{equation}
  When network state $\boldsymbol{s}(t_m)$ at step $t_m$ completely matches state $\boldsymbol{s}(t_n)$ at step $t_n$, the overlap $C(t_m,t_n)$ takes the value $1$.
  When network state $\boldsymbol{s}(t_m)$ is orthogonal to $\boldsymbol{s}(t_n)$, $C(t_m,t_n)$ takes the value $0$.
  Lastly, $C(t_m,t_n)$ takes the value $-1$ when state $\boldsymbol{s}(t_m)$ at step $t_m$ is the reverse of state $\boldsymbol{s}(t_n)$ at step $t_n$, i.e., when $\forall i\in\{1,2,\ldots,N\},2s_i(t_m)-1=-(2s_i(t_n)-1)$.
  
  Figures \ref{fig:phdiag}(b), \ref{fig:phdiag}(c), and \ref{fig:phdiag}(d) show $C(t_m,t_n)$ as heat maps representing the results for different parameter combinations.
  Note that orthogonal elements take the value $1$ since they denote $C(t_n,t_n)$.
  Figure \ref{fig:phdiag}(b) shows the case with $\tau=5$ and $U_{SE}=0.05$,
  where circular motion is not observed.
  Because the values of $C(t_m,t_n)$ are almost $1$, the network converges.
  Figure \ref{fig:phdiag}(c) shows the case with $\tau=20$ and $U_{SE}=0.1$,
  where the circular motion is observed.
  In this case, periodic behavior occurs.
  Finally, Fig. \ref{fig:phdiag}(d) shows the case with $\tau=50$ and $U_{SE}=0.5$,  
  where circular motion is not observed.  
  Here, the overlaps $C(t_m,t_n)$ take values around $0$ except for orthogonal elements, meaning that the network moves randomly.
 
  Here, we summarize the observed results.
  First, spurious states in a region where the synaptic depression is weak remain in equilibrium.
  Second, circular motion begins with the synaptic depression becomes stronger.
  Third, the network moves randomly in a region of very strong synaptic depression.

  \section{Conclusion}
  In this study, we investigated how synaptic depression affects network behavior in memory states and spurious states by using Monte Carlo simulation.
  Since both states are equilibrium, there is little difference between their dynamics.
  The associative memory model is defective as an information processing model because of this indistinguishability.

  We first investigated the dependence of overlaps between the memory pattern and network state on step, in the cases with and without synaptic depression.
  As a result, we found that synaptic depression does not affect memory states but destabilizes spurious states and induces periodic oscillation.
  The period of oscillation decreases as the degree of synaptic depression strengthens.
  These results show that synaptic depression provides the difference in dynamics between memory states and spurious states.
   
  Next, by using PCA which is widely used technique for dimensionality reduction, we found that the network oscillations induced by synaptic depression are explained by circular motion on a plane
  given by the first and second principal components, regardless of initial states.
  Furthermore, we found that the oscillation mainly occurs on subspace spanned by eigenvectors of synaptic weight matrix.
 
  We also obtained a phase diagram showing how the network dynamics depends on the strength of synaptic depression.
  A network is in equilibrium when the synaptic depression is weak.
  As the degree of synaptic depression becomes stronger, the network begins to oscillate circularly.
  Eventually, when the synaptic depression becomes much stronger, the network moves randomly.

 \begin{acknowledgements}
  This work was partially supported by the Aihara Project, part of the FIRST Program from JSPS, initiated by CSTP, and partially supported by JSPS Grants-in-Aid for Scientific Research Grant Numbers 20240020, 22700230, 25106506, 25330283, and 12J06501.
 \end{acknowledgements}
  \bibliographystyle{jpsj}
  \bibliography{jpsj_en_131024}
  \end{document}